\documentclass[iicol,sn-mathphys]{sn-jnl}

\jyear{2021}%

\theoremstyle{thmstyleone}%
\theoremstyle{thmstyletwo}%

\theoremstyle{thmstylethree}%

\begin{document}

\title[ ]{Tuning of Efimov states in non-integer dimensions}


\author*[1]{\fnm{Eduardo} \sur{Garrido}}\email{e.garrido@csic.es}

\author[2]{\fnm{Aksel S.} \sur{Jensen}}\email{asj@phys.au.dk}
\equalcont{These authors contributed equally to this work.}

\affil*[1]{\orgdiv{Instituto de Estructura de la Materia}, \orgname{IEM-CSIC}, \orgaddress{\street{Serrano 123}, \postcode{E-28006} \city{Madrid}, \country{Spain}}}

\affil[2]{\orgdiv{Department of Physics and Astronomy}, \orgname{Aarhus University}, \orgaddress{\postcode{DK-8000} \city{Aarhus C}, \country{Denmark}}}


\abstract{
The purpose of this paper is to show that, by combining Feshbach resonances 
with external confining potentials, the energy scale factor of neighboring 
Efimov states can be tremendously reduced. The Efimov conditions can 
be reached for systems made of three different particles. For the case
of two identical light particles and a heavy particle the energy factor
can be reduced by many orders of magnitude, and the Efimov states are in
this way more easily reachable experimentally.
The equivalence between external potentials and the formulation in
terms of non-integer dimensions, $d$, is exploited.  The technically
simpler $d$-method is used to derive analytic expressions for
two-component relative wave functions describing two short-range
square-well interacting particles.  The two components express one
open and one closed channel. The scattering length is obtained after 
phase shift expansion, providing an analytic form for the Efimov condition. 
We illustrate the results
by means of systems made of $^7$Li, $^{39}$K, and $^{87}$Rb, with
realistic parameters.  The related pairs of dimension and magnetic
field are shown and discussed. The results are universal as they only rely on 
large-distance properties.}




\maketitle

\section{Introduction}

The ability to induce the population of a Feshbach resonance by means of a static magnetic field \cite{ino98,chi10} 
amounts in practice to the possibility of tuning the atom-atom interaction, and, for instance, control the scattering length of the corresponding two-body potential. This fact opened the door to the investigation of three-body systems where at least
one of the internal two-body interactions, in particular, can be tuned to have an infinite scattering length. In this
way conditions for the appearance of the Efimov effect \cite{efi70}
can be satisfied for systems containing three particles. The case of the Cesium trimers \cite{hua14} or the Li-Cs-Cs system \cite{tun14} are good examples of this.

More recently, in Ref.~\cite{gar21}, it has been shown that the Efimov effect can also be induced by an external field that
confines the system. This is based on the fact that, in three dimensions, 
an attractive potential might not bind a two-body system, but in two dimensions the two-body system will always be
bound. As a consequence, if a two-body system is not bound in three dimensions, during the confinement process
to two dimensions, there must necessarily be a moment where the
two-body binding energy is equal to zero, or, in other words, the two-body scattering length is equal to infinity. 
Under these conditions, a three-body system made of two or three identical particles would as well fulfill
the Efimov conditions.

An efficient way to describe the confinement by means of an external field is the $d$-method recently presented
in Ref.~\cite{gar19a}. The idea behind the method is to introduce a continuous dimension $d$, such that the $d$-dependent
centrifugal barrier accounts for the effect of the external field, which does not enter explicitly in the calculation.
The confinement induced Efimov effect was introduced in \cite{nis09}, although the continuous variation of the
squeezing is not investigated.

An external potential along one axis constrains the particles to move
only within a limited volume.  The two extremes would
be a spherical shape and a pancake corresponding respectively to
dimensions $d=3$ and $d=2$.  The equivalence to a dimension dependent
centrifugal barrier has been demonstrated in Refs.~\cite{gar19b,gar20}, where the
two limits, $d=2,3$, are rigorously proved.
The $d$-wave function is then interpreted in the ordinary three-dimensional space as a 
wave function deformed along the confinement axis. In particular, when the Efimov
states are present, they must then be elongated following the
deformed external field, i.e., perpendicularly to the squeezing direction.

The equivalence between a non-integer dimension $d$ and the length, $b_{ho}$, of
an external harmonic oscillator squeezing potential can be found in \cite{gar19b}
for two-body systems. For three-body systems with identical particles, assuming
harmonic oscillator particle-particle interactions with harmonic oscillator length
$r_0$, the connection between $d$ and $b_{ho}$ was derived to be \cite{gar20}
\begin{equation}
\frac{b_{ho}}{r_0}=\sqrt{\frac{2}{3}} \frac{b_{ho}}{r_\mathrm{2D}}
=\sqrt{\frac{2(d-2)}{(d-1)(3-d)} },
\label{eq1}
\end{equation}
where $r_\mathrm{2D}$ is the rms radius of the three-body system in two dimensions
with the realistic interactions.

The ``$d$ vs. $b_{ho}/r_\mathrm{2D}$'' curve in Eq.(\ref{eq1}) is to a large extent universal, 
valid not only for particle-particle harmonic oscillator potentials in the three-body system.
This is shown in Ref.~\cite{gar20}, where the cases of Gaussian and Morse potentials
are considered. When different species are involved, the intermediate $d$-values may correspond to 
different external fields. However, still correct in both limits, Eq.(\ref{eq1}) must be semi-quantitatively, 
or at least qualitatively, correct for all $d$ in the interval, $2<d<3$.  In any case fine tuning 
would probably be necessary for any specific experiment.

The confinement length, $b_{ho}$, is related to the harmonic frequency, $\omega$, by the relation
$\omega=\hbar/(m b_{ho}^2)$, where $m$ is the mass of the atom. The currently available confinement
frequencies move in the range of hundreds of Hz, which for a medium size atom implies that 
$b_{ho}/r_0$ should even approach 1000. This value amounts to a very weak confinement, 
since, as we can see from Eq.(\ref{eq1}), for $b_{ho}/r_0\approx 1000$ we get
$d\approx 2.999999$, extremely close to 3. Note that, for instance, for $d=2.9$ we have
$b_{ho}/r_0 \approx 3.1$. This leads to $\omega$ in the range of tens of MHz, which in principle is 
nowadays out of experimental reach.

The critical dimension, $d=d_E$, for which the two-body potential has infinite scattering length
describes the confinement scenario at which the corresponding two-body state has zero energy.
When this happens for two of the three subsystems a series of infinitely many three-body bound 
states appears, where ratios of energy and size of neighboring states are constant throughout the series.
This is the hallmark of appearance of the Efimov effect. The properties
and behavior of the Efimov states for arbitrary $d$, and how they appear and disappear
in the vicinity of $d=d_E$ are investigated in \cite{gar21,gar21b}. Under these conditions the 
details of the short-distance two-body potentials play a very minor role, and the physics has
therefore a universal character. For this reason, much of the physics can then be captured with simple 
potentials which could even provide analytic solutions. 
The universal connection between scattering length, magnetic field strength, and the measured molecular binding 
energy is derived analytically in Ref.~\cite{lan09} for $d=3$. Here a two-channel model using finite-range 
square-well potentials avoids the zero-range singularity seen in Ref.~\cite{soe13}. These detailed relations are 
successfully tested in the experimentally known case of cesium molecules. The strong variations around the 
Feshbach resonance positions are very well reproduced.

The purpose of this work is to use a similar two-channel model with square-well potentials, but in $d$ dimensions. 
The goal is to study, for a general, not necessarily integer, dimension $d$, the behavior of the scattering length under the presence of a static magnetic field. We are therefore using the $d$-method as theoretical framework.
The model incorporates simultaneously the two different procedures to modify the 
effective two-body scattering length, the dimension $d$ (or the confinement parameter) and the magnetic 
field. The presence of these two parameters will allow to tune simultaneously the scattering length for 
two different two-body systems. In this way it could then be possible to simultaneously tune the two interactions
involved in a three-body system containing two identical particles, or even induce the Efimov effect in 
three-body systems with three different constituents. The connection to specific experiments is not in focus in the 
present paper. Although the results are overall applicable, direct experimental comparison needs detailed consideration.

The paper is organized as follows. We derive in Section~\ref{sect2} the analytical
solution of the Schr\"{o}dinger equation for the two-component problem (one open channel and one closed
channel) in $d$ dimensions and square-well
potentials. Making use of the derived wave functions, we, in Section~\ref{sect3}, provide expressions for
the two-component scattering length and effective range. The behavior of the scattering length as a function 
of the dimension and external magnetic field is investigated in Section~\ref{sect4}, where we focus 
on the condition such that the scattering length is infinite.

The possibility of tuning the two-body scattering
length by means of both the dimension and the magnetic field is now exploited to investigate three-body systems 
and the simultaneous tuning of two of the internal two-body potentials. In Section~\ref{sect5} we discuss the
most convenient systems to be investigated when two of the three constituents are identical and when all the
three are different. The specific cases of the $^{87}$Rb-$^7$Li-$^7$Li and $^7$Li-$^{39}$K-$^{87}$Rb
trimers are considered in Sections~\ref{sect6} and \ref{sect7}, respectively. For each of them we determine
the dimension $d$ (related to the length of the squeezing harmonic oscillator potential through Eq.(\ref{eq1}))
and the magnetic field that gives rise to the appearance of the Efimov effect on these systems, and such that
the energy scale factor is relatively small. We finish in Section~\ref{sect8} with a summary and the conclusions. 
Finally, a number of mathematical derivations are collected in appendices. 

\section{Formulation}
\label{sect2}

We consider a two-body system where only two channels, the ground state and one excited state, contribute to its
wave function. We assume that the energy of the system, $E$, is below the excitation threshold, $E^*$,
and refer to the ground state and the excited state channels as open and closed channels, respectively. 

\begin{figure}
\includegraphics[width=1\linewidth]{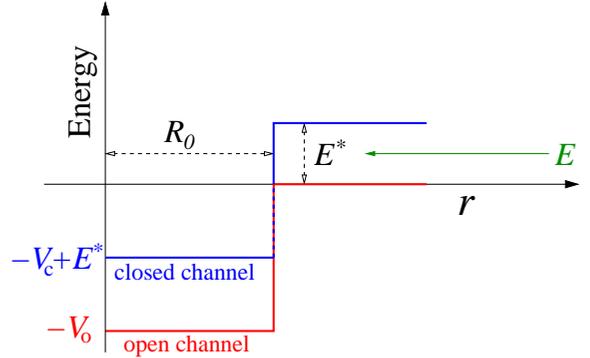}
\caption{Scheme of the square-well potentials for the open and closed channels. The
energies $E$ and $E^*$ are the incident and excitation energies, respectively. Both
square-well potentials have the same radius, $R_0$.}
\label{scheme}
\end{figure}

The open and closed potentials are pictured in Fig.~\ref{scheme}
for the case of square-well potentials with radius $R_0$. The open and closed channels
are coupled through a short-range coupling potential. The effect of the direct scattering process in the open
channel, without coupling to any closed channel, gives rise to the background part of the 
two-body scattering length \cite{mar04}, usually denoted as $a_\mathrm{bg}$.

Below, we derive the analytic expressions for the wave functions and phase shifts of the scattering process 
assuming a relative
$s$-wave state, square-well two-body interactions, and a general, not necessarily integer, dimension $d$.

\subsection{Equations of motion}

The reduced coupled radial Schr\"{o}dinger equations describing the relative
motion in a two-body system with relative energy $E$, and where
only one open and one closed channels contribute, are given by:
\begin{eqnarray} 
\lefteqn{  \hspace*{-0.5cm} 
  \left[ - \frac{\hbar^2}{2\mu} \bigg(\frac{\partial^2}{\partial r^2}
    - \frac{\ell(\ell+1)}{r^2} \bigg) + V_c(r) \right] u_c } \label{e40}   \\ 
    & & \hspace*{2.7cm}  + V_{co} u_o  =  (E-E^*) u_c,  \nonumber \\
\lefteqn{  \hspace*{-0.5cm}
  \left[ - \frac{\hbar^2}{2\mu} \bigg(\frac{\partial^2}{\partial r^2}
    - \frac{\ell(\ell+1)}{r^2} \bigg) + V_o(r) \right] u_o  } \nonumber \\ 
  & & \hspace*{3.7cm}   + V_{oc} u_c = E u_o ,  \label{e45}
\end{eqnarray}
where $u_c$ and $u_o$ are the two components describing the closed and
open channels, respectively, related to the total wave function,
$\Psi$, by
\begin{eqnarray} \label{e50}
 r^{\frac{d-1}{2}} \Psi(r) =  \left( \begin{array}{c}
 u_c(r) \\ u_o(r)
 \end{array} \right).
\end{eqnarray}

In Eqs.(\ref{e40}) and (\ref{e45}) $r$ is the relative coordinate,
$\mu$ is the reduced mass, $V_c(r)$, $V_o(r)$ and $V_{oc}(r)=
V_{co}(r)$ are the diagonal (closed and open) and coupling potentials, $E$ is the energy of
the system, and $E^*$ is the excitation energy from the ground to the excited
state.  The generalized $s$-wave angular momentum, $\ell$, for two
particles depends on the dimension, $d$, and is given by
\cite{gar19a,gar19b,gar20,gar21,nie01}
\begin{eqnarray} \label{e60}
 \ell= \ell_{d,N=2} = \frac{d-3}{2} \;,
\end{eqnarray}
such that the centrifugal barrier strength,
\begin{equation}
\ell(\ell+1)=\frac{(d-3)(d-1)}{4},
\end{equation}
is zero for $d=3$ and $d=1$ and negative and equal to $-1/4$ for
$d=2$. 

In this work we shall assume that all the interactions, $V(r)$, are spherical 
square-well potentials of the same radius, $R_0$, that is of the form (see Fig.~\ref{scheme}):
\begin{eqnarray} \label{e70}
    V(r) = \begin{cases}  - V_0 & \ r < R_0 \\
     \ 0 & \ r > R_0 
    \end{cases},
\end{eqnarray}
where $V_0 (>0)$ represents the different strengths of $V_c(r)$, $V_o(r)$, and
$V_{oc}(r)= V_{co}(r)$, that we will denote, respectively, as $V_c$, $V_o$, and
$V_{co}$.

By introducing the dimensionless coordinate,
$x=r/R_0$, and multiplying Eqs.(\ref{e40}) and (\ref{e45}) by $-2\mu
R^2_0/\hbar^2$,  the equations of motion become
\begin{eqnarray} \label{e80}
 \left( \begin{array}{c}
   u_c^{\prime\prime}(x)  \\
   u_o^{\prime\prime}(x)
 \end{array} \right) = A
 \left( \begin{array}{c}
   u_c(x) \\ u_o(x)
 \end{array} \right) \; ,
\end{eqnarray}
where primes denote derivative with respect to $x$, and the matrix $A$ is defined as
\begin{eqnarray} \label{e87}
 A = \left( \begin{array}{cc}
 \frac{\ell(\ell+1)}{x^2}  - a_+ & -a_{co} \\
 -a_{co}   & \frac{\ell(\ell+1)}{x^2}  - a_-
 \end{array}  \right) \; ,
\end{eqnarray}
where 
\begin{equation}\label{e89}
a_+ = S^2_c - \kappa_c^2, \hspace*{1cm} a_-= S^2_o - \kappa_o^2,
\end{equation}
and where we have used the abbreviations
\begin{equation} \label{e90}
  S^2_c = \frac{2\mu R_o^2 V_c}{\hbar^2}, 
  S^2_o = \frac{2\mu R_o^2 V_o}{\hbar^2},
  a_{co} = \frac{2\mu R_o^2 V_{co}}{\hbar^2},
\end{equation}
all of them positive, and
\begin{eqnarray} \label{e95}
  \kappa_{o} &=& \sqrt{\frac{-2\mu R_o^2 E}{\hbar^2}} =i k_{o} \\ 
  \kappa_{c} &=& \sqrt{\frac{2\mu R_o^2 (E^*-E)}{\hbar^2}}  \label{e97}.
\end{eqnarray} 

Note that with these definitions, $a_+$, $a_-$, $S_c$, $S_o$, $a_{co}$,
$\kappa_o$, $k_o$, and $\kappa_c$, all are dimensionless 
quantities.

\subsection{Solutions}

The square-well potentials require solutions in the two regions, that
is inside and outside the box.  The equations \emph{outside} the box,
$r >R_0 \; (x>1)$, decouple, since then $V_c(r)=V_o(r)=V_{co}(r)=0$, and the matrix
$A$ is diagonal and given by
\begin{eqnarray} \label{e100}
 A^\mathrm{(out)} = \left( \begin{array}{cc}
 \frac{\ell(\ell+1)}{x^2}  +  \kappa^2_{c} & 0 \\
 0   & \frac{\ell(\ell+1)}{x^2}  +  \kappa^2_{o}
 \end{array}  \right) \; .
\end{eqnarray}

With $A=A^\mathrm{(out)}$ in Eq.(\ref{e80}), the two-component solutions are
explicitly for positive energy, $E>0$,
\begin{eqnarray} \label{e110}
& & \hspace*{-1cm} u_c^\mathrm{(out)} = B_c x h_\ell^{(+)}(i\kappa_{c}x) \;, \\ \label{e115}
& & \hspace*{-1cm} u_o^\mathrm{(out)} = B_o x \big(\cos\delta j_\ell(k_{o} x) -
 \sin\delta \eta_\ell(k_{o}x) \big)  \; , 
\end{eqnarray}
where $h_\ell^{(+)}$ is the outgoing Hankel function of first kind, $j_\ell$ and $\eta_\ell$
are, respectively the regular and irregular spherical Bessel functions, and $B_c$ and $B_o$ are normalization constants.
The wave function describing the closed channel falls
off exponentially, and the wave function for the open channel oscillates 
for large distances.  For a negative energy, $E<0$,
also $u_o^\mathrm{(out)}$ must fall off exponentially,
which corresponds to an imaginary $k_{o}$ in Eq.(\ref{e115}).

The equations \emph{inside} the box, $r<R_0 \; (x<1)$, are coupled in
Eq.(\ref{e80}) through the potentials in the non-diagonal general
matrix $A$ in Eq.(\ref{e87}).  The procedure to find the solutions
is to diagonalize this matrix, $A$. The coordinate dependent centrifugal
barrier terms are identical in the diagonal. This fortunately implies
that the transformation matrix, $U$, is a constant independent of the
coordinate, $x$.  We define the resulting diagonal matrix, $D$, as:
\begin{equation} \label{e120}
 U^{-1} A U = D = \left( \begin{array}{cc}
   \frac{\ell(\ell+1)}{x^2} - D^2_+ & 0 \\ 0 & \frac{\ell(\ell+1)}{x^2} - D^2_-
 \end{array}  \right) \; ,
\end{equation}
formed by the eigenvalues of $A$, and where:
\begin{equation} \label{e140}
  D^2_{\pm}  = \frac{1}{2}\bigg( (a_++a_-) \pm
    \sqrt{(a_+-a_-)^2 + 4a_{co}^2}\bigg) \;,
\end{equation}
if $a_+>a_-$, or interchange of the two eigenvalues in the opposite
case of $a_+<a_-$, i.e.,
\begin{equation} \label{e141}
  D^2_{\pm}  = \frac{1}{2}\bigg( (a_++a_-) \mp
    \sqrt{(a_+-a_-)^2 + 4a_{co}^2}\bigg) \;.
\end{equation}

These definitions are chosen such that $D_{\pm}^2
\rightarrow a_{\pm}$ when $a_{co} \rightarrow 0$, independent of the
relative size of $a_+$ and $a_-$. With these definitions we also have that
\begin{eqnarray} \label{e151}
D^2_+ - a_+ = - (D^2_- - a_-) \;, \\ \label{e152}
D^2_+ - a_- =  - (D^2_- - a_+) \;, \\ \label{e153}
(D^2_+ - a_+)(D^2_+ - a_- ) = a_{co}^2 \;.
\end{eqnarray}

The eigenvectors associated to the above eigenvalues
permit to construct the transformation matrix,
$U=U^{-1}$, with normalized column vectors, as:
\begin{eqnarray} \label{e154}
  U = \frac{1}{\sqrt{N}} \left( \begin{array}{cc}
      D^2_+ - a_-  &  a_{co}    \\
     a_{co} &   D^2_- - a_+  
 \end{array}  \right) \; ,
\end{eqnarray}
where the normalization is given by 
\begin{eqnarray} \label{e156}
  N &=& (D^2_+ - a_-)^2 +  a_{co}^2 = (D^2_- - a_+)^2 +  a_{co}^2
  \nonumber \\  &=& (D^2_+ - a_-) (D^2_+ - D^2_-)  \;.
\end{eqnarray}
The difference and sum of $a_+$ and $a_-$, which enter in the definitions of
$D_{\pm}$ in Eqs.(\ref{e140}) and (\ref{e141}), are given by
\begin{equation} \label{e320}
  a_+ - a_- = S^2_c -S^2_o - \frac{2\mu R_0^2 E^*}{\hbar^2}  \;.
\end{equation}
\begin{equation} \label{e322}
a_+ + a_- = S^2_c+S^2_o - \frac{2\mu R_0^2 E^*}{\hbar^2}+ \frac{4\mu R_0^2 E}{\hbar^2}\;.
\end{equation}

The diagonal problem:
\begin{equation}
 \left( \begin{array}{c}
 \phi_+^{\prime\prime} \\ \phi_-^{\prime\prime}
  \end{array} \right)= D  \left( \begin{array}{c}
 \phi_+ \\ \phi_-
  \end{array} \right)
  \label{eqd}
\end{equation}
leads to the uncoupled equations for $\phi_{\pm}$:
\begin{eqnarray} \label{e134}
  \phi_{\pm}^{\prime\prime} = \bigg(\frac{\ell(\ell+1)}{x^2} - D^2_{\pm} \bigg) \phi_{\pm} \;,
\end{eqnarray}
whose solutions, regular at the origin, are
\begin{eqnarray} \label{e130}
  \phi_{\pm} = C_{\pm} x j_\ell(xD_{\pm}) \; ,
\end{eqnarray}
where $C_{\pm}$ are normalization constants and $j_\ell$ is the
regular spherical Bessel function.

Combining now Eqs.(\ref{e80}) and (\ref{e120}) it is simple to see that:
\begin{equation}
U^{-1} \left( \begin{array}{c}
   u_c^{\prime\prime}(x)  \\
   u_o^{\prime\prime}(x)
 \end{array} \right) = D U^{-1}
 \left( \begin{array}{c}
   u_c(x) \\ u_o(x)
 \end{array} \right),
\end{equation}
which, by comparison to Eq.(\ref{eqd}), immediately leads to:
\begin{eqnarray} \label{e125}
  \left( \begin{array}{c}
 u_c^\mathrm{(in)} \\ u_o^\mathrm{(in)}
  \end{array} \right) = U 
 \left( \begin{array}{c}
 \phi_+ \\ \phi_-
  \end{array} \right)  \;,
\end{eqnarray}
which, by combination of Eqs.(\ref{e154}) and (\ref{e130}), permit to obtain
the solutions inside the box:
\begin{equation} \label{e250}
  u_c^\mathrm{(in)} =  (D^2_+-a_-) C_+ x j_\ell(xD_+) + a_{co} C_- x j_\ell(xD_-)
\end{equation}
\begin{equation} \label{e255} 
  u_o^\mathrm{(in)} =  (D^2_--a_+) C_- x j_\ell(xD_-) + a_{co} C_+ x j_\ell(xD_+), 
\end{equation}
where we have omitted the unimportant normalization constant $N$.

\section{Scattering length and effective range}
\label{sect3}

The matching of the logarithmic derivatives of the solutions inside and
outside the box, i.e., the matching of Eqs.(\ref{e110}) and (\ref{e250}), 
and Eqs.(\ref{e115}) and (\ref{e255}), at $x=1$ is described in 
Appendix~\ref{appmatch}. This leaves to the following
expression for the phase shift, Eq.(\ref{e443ap}):
\begin{eqnarray} \label{e443}
  \cot\delta =  \frac{\eta_\ell(k_{o})}{j_\ell(k_{o})}
  \frac{F_2  - k_{o} \frac{\eta_{\ell+1}(k_{o})}{\eta_\ell(k_{o})}}
  {F_2 - k_{o} \frac{j_{\ell+1}(k_{o})}{j_\ell(k_{o})}} \; ,
\end{eqnarray}
where $F_2$ is given by Eq.(\ref{e456}).

The phase shift can be expanded in powers of energy, resulting in the
effective range expansion defining the scattering length and the effective
range.  The lowest order term is then obtained by expanding in energy
up to first order in $E$, or second order in $k_{o}$, and inserting
energy zero ($E=0$) in all resulting terms.  The exception is the
overall energy factor arising from $\eta_\ell(k_{o})$ and
$j_\ell(k_{o})$, which however also must be expanded to include the
two lowest orders in energy.

For the sake of clarity, we give the details of the derivations 
in Appendix~\ref{apareff}, and here we only quote and discuss the final results.

\subsection{Expanding the phase shift}

After the low-energy expansion of all the terms in Eq.(\ref{e443})
we get that, Eq.(\ref{e666}):
\begin{eqnarray} \label{delexp}
 && \frac{(2\ell+1)}{[(2\ell+1)!!]^2} k_o^{2\ell+1}  \cot\delta \simeq
  -\bigg(1 - \frac{2\ell +1}{F_2^{(0)}}\bigg)
  \\ \nonumber &&
   - E \,\, \frac{2\ell+1}{F_2^{(0)}}\bigg[
  \frac{F_2^{\prime(0)}}{F_2^{(0)}} - \frac{k_o^2}{E} \frac{1}{2\ell+3}
  \left(1+\frac{1}{F_2^{(0)}}-\frac{F_2^{(0)}}{2\ell-1} \right) 
  \bigg],
\end{eqnarray}
where for non-integer values of $\ell$ the factor $(2\ell+1)!!$
has to be replaced by $\Gamma(2\ell+2)/(2^\ell\Gamma(\ell+1))$, 
$F_2^{(0)}$ is given by Eq.(\ref{e577}),
$F_2^{\prime(0)}=\partial F_2^{(0)}/\partial E$, and the superscript $(0)$,
wherever it appears, refers from now on to the function evaluated at $E=0$. 

The above expression can be compared to 
\begin{eqnarray} \label{e667}
  & &\frac{(2\ell+1)}{[(2\ell+1)!!]^2} k_o^{2\ell+1}  \cot\delta
  \rightarrow - \bigg(\frac{R_0}{a_\mathrm{tw}} -
  \frac{1}{2}\frac{R_\mathrm{eff}}{R_0} k_o^2  \bigg)^{2\ell +1} \\
   & &= -\left(\frac{R_0}{a_\mathrm{tw}} \right)^{2\ell+1}  \!\!\!\!\!\!\!\! +(2\ell+1)
   \left(\frac{R_0}{a_\mathrm{tw}} \right)^{2\ell}
   \frac{1}{2}\frac{R_\mathrm{eff}}{R_0} k_o^2+{\cal O}(k_o^4), \nonumber
\end{eqnarray}
where $a_\mathrm{tw}$ and $R_\mathrm{eff}$ are the two-component scattering
length and the effective range, respectively.

From Eq.(\ref{e667}), after comparison with Eq.(\ref{delexp}), and
keeping in mind that $E= \hbar^2 k_0^2/(2\mu R^2_0)$, we can 
identify:
\begin{equation} \label{e673}
  \frac{a_\mathrm{tw}}{R_0}  =  \left( 1-\frac{2\ell+1}{F_2^{(0)}} \right)^{-\frac{1}{2\ell+1}}.
\end{equation}

If $a_{co}=0$, $a_\mathrm{tw}=a_\mathrm{bg}$, and Eq.(\ref{e673}) reduces to: 
\begin{equation} 
  \frac{a_\mathrm{bg}}{R_0}  =  
  \left( 1-(2\ell+1)\frac{ j_\ell(S_o)}{S_o j_{\ell+1}(S_o)} \right)^{-\frac{1}{2\ell+1}},
\label{anocou}
\end{equation}
which is the expression of the scattering length for the square-well potential and an 
arbitrary relative partial wave, $\ell$.

From Eqs.(\ref{delexp}) and (\ref{e667}) we can also extract the effective range as
shown in Appendix~\ref{appc}.

\subsection{Connection to open and closed channels}

Eq.(\ref{e673}) gives the two-component scattering length for square-well potentials. To make it more
general, it is convenient to write it in terms of the scattering lengths for the closed and open channels, 
$a_\mathrm{closed}$ and $a_\mathrm{open}$. With this in mind, we define:
\begin{equation}
\left(\frac{a_\mathrm{closed}}{R_0}\right)^{2\ell+1}=\left[
1-(2\ell+1)\frac{j_\ell(D_+^{(0)})}{D_+^{(0)}j_{\ell+1}(D_+^{(0)})} 
\right]^{-1},
\label{acl}
\end{equation}
and
\begin{equation}
\left(\frac{a_\mathrm{open}}{R_0}\right)^{2\ell+1}=\left[
1-(2\ell+1)\frac{j_\ell(D_-^{(0)})}{D_-^{(0)}j_{\ell+1}(D_-^{(0)})} 
\right]^{-1},
\label{aop}
\end{equation}
such that in the limit of no coupling, $a_{co}=0$,
we recover the known ${\ell}$-dependent square-well expression,
Eq.(\ref{anocou}), and therefore $a_\mathrm{open}=a_\mathrm{bg}$.

In the same way we define $\tilde{\kappa}_c$ as:
\begin{equation}
 \frac{1}{\tilde{\kappa}_c^{ 2\ell+1}}=\left[
 1-(2\ell+1)\frac{K_{\ell+\frac{1}{2}}(\kappa_c^{(0)})}{\kappa_c^{(0)}K_{\ell+\frac{3}{2}}(\kappa_c^{(0)})}
 \right]^{-1},
\label{eq47}
\end{equation}
which, for any $\ell$, in the presence of a very weakly bound
state becomes $\tilde{\kappa}_c \rightarrow R_0/a_\mathrm{closed}$.

Using these definitions, it is not difficult to write $F_2^{(0)}$ in Eq.(\ref{e577}) as
a function of $a_\mathrm{closed}$, $a_\mathrm{open}$, and $\tilde{\kappa}_c$, and 
get that the scattering length in Eq.(\ref{e673}) becomes: 
\begin{eqnarray}
\lefteqn{\hspace*{-5mm}
\frac{(2\ell+1)}{[(2\ell+1)!!]^2} k_o^{2\ell+1}  \cot\delta \rightarrow
-\left( \frac{R_0}{a_\mathrm{tw}}\right)^{2\ell+1}\!\!\!\!\!\!= 
-\left(\frac{R_0}{a_\mathrm{open}}\right)^{2\ell+1}   \!\!\!\!\!\! 
} \nonumber \\ && \hspace*{-8mm}+
\frac{ \left(\frac{R_0^{2\ell+1}}{a^{2\ell+1}_\mathrm{open}}-\tilde{\kappa}_c^{2\ell+1} \right)
\left( \frac{R_0^{2\ell+1}}{a^{2\ell+1}_\mathrm{open}} - \frac{R_0^{2\ell+1}}{a^{2\ell+1}_\mathrm{closed}}\right)
\frac{a^2_{co}}{(D_-^{(0)2}-a_+^{(0)})^2}}
{\left(\frac{R_0^{2\ell+1}}{a^{2\ell+1}_\mathrm{closed}}-\tilde{\kappa}_c^{2\ell+1} \right)+\left(\frac{R_0^{2\ell+1}}{a^{2\ell+1}_\mathrm{open}}-\tilde{\kappa}_c^{2\ell+1} \right) \frac{a^2_{co}}{(D_-^{(0)2}-a_+^{(0)})^2} 
}.\nonumber  \\ \label{eq93}
\end{eqnarray}

When $a_{co}\rightarrow 0$, an expansion of Eq.(\ref{eq93}) up to first order in $a_{co}^2$
permits to reduce the equation to:
\begin{eqnarray}
\lefteqn{
\frac{(2\ell+1)}{\left[(2\ell+1)!!\right]^2}k_0^{2\ell+1}  \cot\delta \rightarrow } \label{eq61} \\ && 
-\left( \frac{R_0}{a_\mathrm{tw}}\right)^{2\ell+1}=
-\left(\frac{R_0}{a_\mathrm{open}}\right)^{2\ell+1} \!\!\!\!\!\!\!  + 
\frac{\beta^2}{\frac{R_0^{2\ell+1}}{a_\mathrm{closed}^{2\ell+1}} -  \tilde{\kappa}_c^{2\ell+1}  },
\nonumber
\end{eqnarray}
where the parameter $\beta^2$ is just the numerator in the last term of Eq.(\ref{eq93}), which contains
$a_{co}^2$ multiplied by a series of, either finite or small, factors.

By identifying $\beta$ in Eq.(\ref{eq61}) and the coupling parameter $\beta$ employed in Ref.~\cite{soe13},
we can see that Eq.(\ref{eq61}) reduces trivially for three dimensions ($\ell=0$) to Eq.(7) of Ref.~\cite{soe13}, 
where the two-component problem is investigated in three dimensions with a contact interaction model.

\section{Critical dimension}
\label{sect4}

Equation (\ref{eq93}) gives the scattering length for the two-component problem in $d$ dimensions, from which 
we  write:
\begin{eqnarray}
\lefteqn{
a_\mathrm{tw}^{2\ell+1}=   a^{2\ell+1}_\mathrm{open} } \label{eq25} \\ && \hspace*{-1cm}
\times  \frac{ 
\left(\frac{R_0^{2\ell+1}}{a^{2\ell+1}_\mathrm{closed}}-\tilde{\kappa}_c^{2\ell+1} \right)
+\left(\frac{R_0^{2\ell+1}}{a^{2\ell+1}_\mathrm{open}}-\tilde{\kappa}_c^{2\ell+1} \right) \frac{a^2_{co}}{(D_-^{(0)2}-a_+^{(0)})^2}
}{ 
\left(\frac{R_0^{2\ell+1}}{a^{2\ell+1}_\mathrm{closed}}-\tilde{\kappa}_c^{2\ell+1} \right)
+\frac{a^{2\ell+1}_\mathrm{open} }{a^{2\ell+1}_\mathrm{closed} }
\left(\frac{R_0^{2\ell+1}}{a^{2\ell+1}_\mathrm{open}}-\tilde{\kappa}_c^{2\ell+1} \right) \frac{a^2_{co}}{(D_-^{(0)2}-a_+^{(0)})^2}
}. \nonumber
\end{eqnarray}

It is obvious that the two-component scattering length, $a_\mathrm{tw}$, becomes infinite when 
the denominator of the fraction in the expression above is equal to zero. After some rewriting, this
condition can be written as:
\begin{eqnarray}\label{eq98}
  \tilde{\kappa}_c^{2\ell+1} &=& \frac{R_0^{2\ell+1}}{a^{2\ell+1}_\mathrm{closed}}
 \\  \nonumber &+& \frac{1-a^{2\ell+1}_\mathrm{open}/a^{2\ell+1}_\mathrm{closed} }
{\frac{a^{2\ell+1}_\mathrm{closed}}{R_0^{2\ell+1}}+
\frac{a^{2\ell+1}_\mathrm{open}}{R_0^{2\ell+1}}\frac{a^2_{co}}{(D_-^{(0)2}-a_+^{(0)})^2} }
\frac{a^2_{co}}{(D_-^{(0)2}-a_+^{(0)})^2}.
\end{eqnarray}

Note that the quantities $\tilde{\kappa}_c$ in Eq.(\ref{eq47}),
$a_\mathrm{closed}$ in Eq.(\ref{acl}), and $a_\mathrm{open}$ in
Eq.(\ref{aop}), are, through $\ell$ in Eq.(\ref{e60}), functions of the dimension $d$. We define
the critical dimension, $d=d_E$, as the value of $d$ such that
the equality in Eq.(\ref{eq98}) is fulfilled. For this dimension we have that $a_\mathrm{tw}=\infty$,
or, equivalently, the two-body two-component system has zero binding energy.

The momentum, $\tilde{\kappa}_c$ in Eq.(\ref{eq47}), is a function of $d$ and $\kappa_c^{(0)}$, 
which is given in Eq.(\ref{e372}). Also, $a_+^{(0)}$ in Eq.(\ref{e373}), and therefore 
$D_\pm^{(0)}$ in Eq.(\ref{e371}), contain as well a dependence on $\kappa_c^{(0)}$. If the system is under the influence 
of an external static magnetic field, $B$, the momentum $\kappa_c^{(0)}$ then becomes:
\begin{equation}
\kappa_c^{(0)}\rightarrow \kappa_c^{(0)}(B) = \sqrt{\frac{2\mu R_0^2 (E^*-\delta_\mu B)}{\hbar^2}},
\label{extb}
\end{equation}
where $\delta_\mu$ is the difference between the magnetic moments of the system in the ground (open)  
and the excited (closed) states.

Therefore, the condition in Eq.(\ref{eq98}) that determines when $a_\mathrm{tw}=\infty$,
is now a function of $d$ and $B$. In other words, one could get the curve ``$B_0$ vs. $d_E$'' 
formed by the $(B_0,d_E)$ points satisfying  Eq.(\ref{eq98}).

\section{Properties for three-body systems}  
\label{sect5}

Let us consider a three-body system formed by two identical particles and a 
different one. This system  involves two different two-body interactions. If we 
plot the ``$B_0$ vs. $d_E$'' curve for each of these two interactions, and they happen to
cross, we then have the dimension $d=d_E$ and magnetic field $B=B_0$, that
make $a_\mathrm{tw}=\infty$ simultaneously for both potentials. In this way it would be possible
to get the Efimov condition fulfilled for all the two-body subsystems. 
Furthermore, following the same procedure, the Efimov conditions can also be fulfilled
in a three-body system made of three different constituents, since with this method it is, 
in principle, possible
to tune $a_\mathrm{tw}=\infty$ for two of the three two-body interactions involved.

Let us then focus on these two different three-body systems, one made of two identical particles and
one different (ABB system), and a system made of three different particles (ABC system).  Here
we analyze the scaling between consecutive Efimov states, and how to construct the appropriate two-body potentials.

\subsection{Mass dependence of the scale factor}

\begin{figure}
\includegraphics[width=1\linewidth]{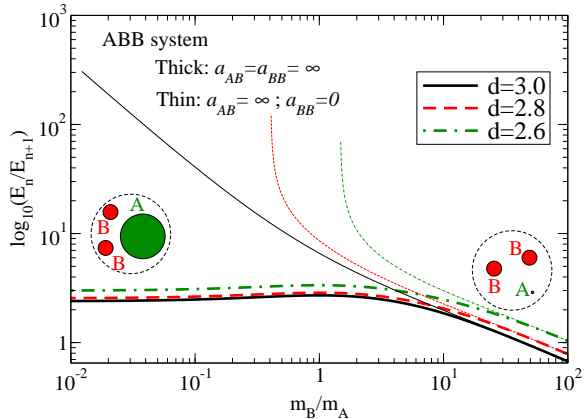}
\caption{Logarithm of the energy scale factor between two consecutive Efimov states for an ABB system, as a function of 
the $m_B/m_A$ mass ratio. The scattering length between particles A and B, $a_{AB}$, is equal to $\infty$. The thick
and thin lines show the results when the scattering length between the two identical particles, $a_{BB}$ is equal
to $\infty$ or 0, respectively. The results for three different dimensions, $d=3.0$ (solid), 2.8 (dashed), and 2.6
(dot-dashed), are shown. }
\label{fig0}
\end{figure}

Let us recall that for the ABB case, when only the AB scattering length is sufficiently large, 
the situation of two heavy particles and one light particle is particularly favorable for the appearance of 
Efimov states \cite{gar18}. 
In this case, the energy scale factor between two consecutive states is significantly reduced. As a consequence, its experimental determination, or even its numerical calculation,  is much easier than in 
the case of two light particles and one heavy particle.

The energy scale factor between two consecutive states is given by 
$E_n/E_{n+1}=\exp{(2\pi/\vert \nu_\infty \vert)}$, where $E_n$ is the binding energy of the $n^\mathrm{th}$ 
Efimov state and $\vert E_n\vert > \vert E_{n+1}\vert$ ($n=1,2,3,\cdots$).
The $d$-dependent  $\vert \nu_\infty\vert$ is independent of the interactions and independent of $n$. The value
of  $\vert \nu_\infty\vert$ can be obtained 
through a transcendental equation as shown in Eqs.(26) and (27) of Ref.~\cite{fed01} for $d=3$, and generalized
for arbitrary dimension in subsection 5.2 of Ref.~\cite{nie01}.

The results for the ABB system are shown in Fig.~\ref{fig0}, where we show the logarithm of the energy ratios
between two consecutive Efimov states as functions of the mass ratio $m_B/m_A$. More precisely, the $y$-axis
gives the exponent $a$ when the energy ratio is written as $10^a$.

We first focus on the results when the scattering length between the two identical particles, $a_{BB}$,
is equal to 0, and the one between the two different particles, $a_{AB}$, is equal to infinity (thin curves). 
Three different dimensions, $d=3.0$, 2.8, and 2.6, are considered. 
As we can see, when $m_B \gg m_A$, which corresponds
to two heavy particles and one light particle, the energy scale factor is many orders of magnitude smaller than for
 $m_B \ll m_A$, which corresponds to two identical light particles and one heavy particle.  We can also see
that, the smaller the dimension, the sooner the divergence of the
energy scale factor occurs with decreasing  mass ratio, $m_B/m_A$.

When the two scattering lengths, $a_{AB}$ and $a_{BB}$, are equal to infinity we obtain the results shown
by the thick curves in Fig.~\ref{fig0}. For two heavy particles and one light 
particle, we can see that the effect of making $a_{BB}=\infty$ is very small. However, for two light particles 
and one heavy particle, the ratio between two consecutive Efimov states reduces significantly. For the three dimensions shown in the figure the scale factor ranges between 250 and 1000 when $m_B/m_A \ll 1$.  These ratios are 
comparable, or even smaller, than for three identical bosons in three dimensions, where
the energy scale factor is 515.

\begin{figure}
\includegraphics[width=1\linewidth]{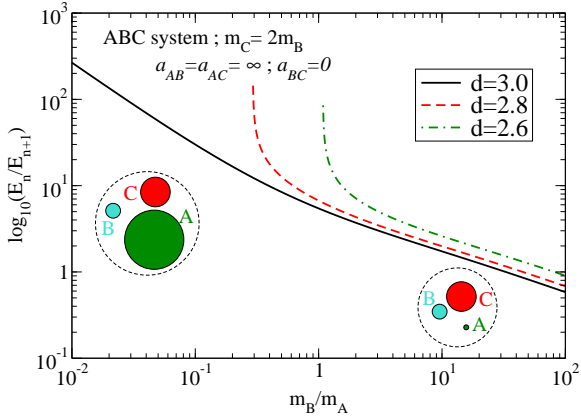}
\caption{The same as Fig.~\ref{fig0} for the ABC system. The mass of particle C is chosen to be twice the mass
of particle B  ($m_C=2 m_B$). The scattering lengths between particles A and B, $a_{AB}$, and between particles 
A and C, $a_{AC}$, are equal to infinity. The scattering length between B and C, $a_{BC}$, is equal to 0.  
The results for three different dimensions, $d=3.0$ (solid), 2.8 (dashed), and 2.6 (dot-dashed), are shown. }
\label{fig0b}
\end{figure}

In Fig.~\ref{fig0b} we show the same as in Fig.~\ref{fig0} but for the ABC system.  The scattering lengths between 
particles A and B, $a_{AB}$, and between particles A and C, $a_{AC}$, are equal to infinity, and the one between B and C, $a_{BC}$, is equal to 0.  The masses of particles
B and C, $m_B$ and $m_C$, have been chosen such that $m_C=2 m_B$. The curves are again shown as a function 
of $m_B/m_A$. The behavior of these curves is very similar to the thin ones
in Fig.~\ref{fig0}. That is, when $m_B \ll m_A$ the system behaves as the case of 
two identical light particles and a heavy particle in Fig.~\ref{fig0}, and the energy scale factor between two consecutive
Efimov levels is enormous. On the contrary, when $m_A$ is much smaller than the other two masses,
the system behaves as in the case of two identical heavy particles and a light particle, with an energy scale factor even
smaller than 10 for $m_B/m_A=100$ and for the three dimensions shown in the figure. The conclusion is then that,
for an ABC system, if one of the particles is much lighter than the other two, and the scattering length 
between this light particle and the other two is sufficiently large, the energy scale factor can be reasonably 
small.

\subsection{Constructing the two-body potentials}

Equations (\ref{eq98}) and (\ref{extb}) permit to obtain the pairs of values, magnetic field and dimension, that
make the two-component scattering length, $a_\mathrm{tw}$, equal to infinity. The magnetic field and dimension
values fulfilling this condition will be denoted as $(B_0,d_E)$.

The goal now is to obtain the $(B_0,d_E)$ points for specific two-body systems. To do so, it will be necessary to specify
the potential used. In gereral, the parameters entering in the interaction will be determined making use of the experimental information available in three dimensions.

The two-component problem in three dimensions has been investigated in Ref.~\cite{soe13}. There, it has
been shown that the two-component scattering length is given by:
\begin{equation}
\frac{1}{a_\mathrm{tw}}=\frac{1}{a_\mathrm{open} }+\frac{\beta^2}{\kappa(B)-a^{-1}_\mathrm{closed}},
\label{atw3d}
\end{equation}
where $\beta^2$ describes the coupling between the open and closed channels, and
\begin{equation}
\kappa(B)=\sqrt{\frac{2\mu}{\hbar^2}(E^*-\delta_\mu B)}.
\label{kapb}
\end{equation}

In Ref.~\cite{soe13}, $a_\mathrm{open}$ is taken equal to $a_\mathrm{bg}$, often known experimentally.
The same equallity, $a_\mathrm{open}=a_\mathrm{bg}$, will be considered here.

From Eq.(\ref{atw3d}), and taking the value of the magnetic field, $B=B_0$, such that
$a_\mathrm{tw}=\infty$, we can easily see that:
\begin{equation}
\frac{1}{a_\mathrm{closed}}=\kappa(B_0)+\beta^2 a_\mathrm{open}.
\label{acl3d}
\end{equation}

In the same way, if we take $B=B_0+\Delta B$ such that $a_\mathrm{tw}=0$, we obtain:
\begin{equation}
\delta_\mu \Delta B=\frac{\hbar^2}{\mu} \kappa(B_0) \beta^2 a_\mathrm{open}.
\label{beta2}
\end{equation}

Feshbach resonances are therefore characterized by the experimental parameters $a_\mathrm{bg}$,
$B_0$, $\Delta B$, and $\delta_\mu$. With this experimental information we can, through Eq.(\ref{beta2}),
determine the coupling term $\beta^2$, and then, through Eq.(\ref{acl3d}), the scattering length for the closed 
channel, $a_\mathrm{closed}$.

The simple two-body square-well interactions used in this work require two parameters to be fully 
determined, the range and the strength. For the range we shall always use the van der Waals length, $R_\mathrm{vdW}$,
often known experimentally. The strength of the open and closed channels, $V_o$ and $V_c$ in Eq.(\ref{e90}), are fixed
in order to reproduce, respectively, the value of $a_\mathrm{open}$, known experimentally, and 
$a_\mathrm{closed}$ as given in Eq.(\ref{acl3d}) for $d=3$. Finally, the coupling strength, $V_{co}$ in
Eq.(\ref{e90}), is obtained by imposing that, for $B=B_0$ and $d=3$, we must get $a_\mathrm{tw}=\infty$.

Even if these constraints permit to determine the range and the strength of the open, closed, and
coupling potentials, there is still some freedom to fix these values. First, the value of $E^*$, which
enters through $\kappa(B_0)$ in Eqs.(\ref{acl3d}) and (\ref{beta2}), is completely arbitrary, although, as
quoted in \cite{soe13}, its value does not affect the final observables significantly. And second, the values
of $V_o$ and $V_c$ can be very different depending on the number of bound states hold by the potentials. 
About $V_{co}$, after fixing $V_o$ and $V_c$, it is still possible to find a family of different values all of them giving rise
to $a_\mathrm{tw}=\infty$.

In any case, in what follows, the goal is not to derive detailed two-body potentials for the
systems under investigation, but specifically to show that, by a combination of the confinement and external 
magnetic field, it is in principle possible to simultaneously tune the two interactions involved in the ABB-systems,
or two out of the three interactions involved in the ABC-systems. In this way, the energy separation between the Efimov states in ABB-systems can be drastically reduced,
and even systems involving three different constituents could exhibit the Efimov effect with a reasonably small energy separation between the states.

\section{The ABB-system}
\label{sect6}

\begin{table*}[ht!]
\begin{center}
\begin{minipage}{\textwidth}
\caption{For the three dimers considered in this work, we give the reference from which the Feshbach resonance
parameters $B_0$, $\Delta B$, $\delta_\mu$, $a_\mathrm{open}$, and $R_\mathrm{vdW}$, also given in the table, are taken. Using
the excitation energy, $E^*=5$ K, the quoted Feshbach resonance parameters give rise to the scattering length for
the closed channel potential, $a_\mathrm{closed}$. Finally, the last three columns give the strengths of the square-well potentials for the
open, closed, and coupling channels such that the given $a_\mathrm{open}$ and $a_\mathrm{closed}$ values are reproduced, and
$a_\mathrm{tw}=\infty$ for $d=3$ and $B=B_0$. The magnetic fields are given in Gauss (G), the energies and potential strengths are given
in Kelvin (K), the distances are given in units of the Bohr radius, $a_0$, and the magnetic moment difference, $\delta_{\mu}$, is given in units of the Bohr magneton, $\mu_B$.}
\label{tab1}
\begin{small}
\begin{tabular}{c|cccccc|c|ccc}
\toprule
dimer &    Ref. & $B_0$ (G)  &  $\Delta B$ (G) & $\delta_\mu (\mu_B)$ &  $a_\mathrm{open} (a_0)$  & $R_\mathrm{vdW} (a_0)$  &  $a_\mathrm{closed} (a_0)$ &  $V_o$ (K)  & $V_c$ (K) & $V_{co}$ (K) \\ \midrule
$^7$Li-$^{87}$Rb & \cite{mar09}  &  649   &  $-70   $  &  2.0   &  $-36$  &  44.1  &  1.65  &  28.448  &  28.441  &  4.773  \\
$^7$Li-$^7$Li &   \cite{chi10}   &  737   &  $-192.3$  &  1.93  &  $-25$  &  33.1  &  2.25  &  20.049  &  20.026  &  3.883  \\
$^7$Li-$^{39}$K &  \cite{cui18} &  319   &  $30$      &  1.5   &  $89$  &  42.1  &   1.71  &  17.072  &  17.040  &  2.505  \\ 
\botrule
\end{tabular}
\end{small}
\end{minipage}
\end{center}
\end{table*}

The energy scale factor for an ABB system was discussed in Fig.~\ref{fig0}. In particular, we
saw that, when particle A is much heavier than the other two, having the two scattering lengths, $a_{AB}$
and $a_{BB}$, simultaneously equal to infinity, enormously reduces the separation between consecutive Efimov states.

To illustrate this fact, we choose the $^{87}$Rb-$^7$Li-$^7$Li system, which in principle should not be a good candidate
to exhibit the Efimov effect. In order to investigate this system it is necessary first to construct the
required $^7$Li-$^{87}$Rb and $^7$Li-$^7$Li two-body potentials.

\subsection{The $^7$Li-$^{87}$Rb dimer}

The experimental information about this system can be found in Ref.~\cite{mar09}. In particular, they give a value for the $C_6$
coefficient in the van der Waals potential of $2550 E_h a_0^6$ ($E_h$ is a Hartree and $a_0$ the Bohr radius), from
which one can get a van der Waals length of $R_\mathrm{vdW}=44.1 a_0$, which is taken as the potential range of our 
square-well potentials. The connection between the $C_6$ coefficient and $R_\mathrm{vdW}$ is given by \cite{chi10}:
\begin{equation}
R_\mathrm{vdW}=\frac{1}{2}\left( \frac{2\mu C_6}{\hbar^2} \right)^{1/4},
\end{equation}
where $\mu$ is the reduced mass of the system.

In \cite{mar09} they also found a broad Feshbach resonance centered at $B_0=649$ G, with $\Delta B=-70$ G, and
$a_\mathrm{bg}=a_\mathrm{open}=-36 a_0$. Although the value for $\delta_\mu$ is not given, we shall use the 
typical value $\delta_\mu = 2\mu_B$ (where $\mu_B$ is the Bohr magneton).

With these values we have that $\delta_\mu B_0=0.087$ K, which implies that $E^*$ must be larger
than this value (otherwise, see Eq.(\ref{kapb}), we would get an imaginary value of $\kappa(B_0)$).
In particular we have chosen $E^*=5$ K, which is sufficiently larger than $\delta_\mu B_0$ such that the 
final results are almost independent of the specific value of $E^*$.
Using the values of $a_\mathrm{open}$, $E^*$, $B_0$, $\Delta B$, and $\delta_\mu$ given above we get from Eq.(\ref{acl3d})
that $a_\mathrm{closed}=1.65 a_0$.

We therefore choose the strengths $V_o$ and $V_c$, such that the values of $a_\mathrm{open}=-36 a_0$
and $a_\mathrm{closed}=1.65 a_0$ are reproduced.
To do so, one could in principle choose the minimum possible strength, which amounts to a two-body potential
not even binding the two-body dimer in the case of negative scattering length. This is very unrealistic, since the number of bound $^7$Li-$^{87}$Rb
states is probably quite high. For this reason we have chosen a strength for the open channel $V_o=28.448$ K,
which gives rise to about 20 bound dimer states. For the closed channel, among all the possible strengths reproducing
the desired value of $a_\mathrm{closed}$, we have taken $V_c= 28.441$ K as the value closest to the chosen $V_o$ value.

For the coupling potential, $V_{co}$, adjusted to obtain $a_\mathrm{tw}=\infty$ for $B=B_0$, we have chosen a value
clearly smaller than $V_o$ and $V_c$, but not small enough to be a perturbation. In particular
we have taken $V_{co}=4.733$ K.

The selected values of these potential parameters are collected in the first row of Table~\ref{tab1}.

\subsection{The $^7$Li-$^{7}$Li dimer}

For the $^7$Li-$^{7}$Li interaction we proceed in the same way using the information provided in Ref.~\cite{chi10}.
The van der Waals length is equal to 33.1 $a_0$, which is taken as the range of the potentials.

In three dimensions, the Feshbach resonance for the $^7$Li-$^{7}$Li system is characterized by
$B_0=737$ G, $\Delta B=-192.3$ G, $a_\mathrm{open}=-25 a_0$, and $\delta_\mu=1.93 \mu_B$. These
values correspond to $\delta_\mu B_0=0.096$ K, and, as for the $^7$Li-$^{87}$Rb system, we choose
$E^*=5$ K. 

With all this information, by means of Eq.(\ref{acl3d}), we now get $a_\mathrm{closed}=2.25 a_0$.
The strengths of the open and closed channels have been chosen to be $V_o=20.049$ K and
$V_c=20.026$ K, which reproduce, respectively, the desired values of $a_\mathrm{open}$ and
$a_\mathrm{closed}$. With these strengths the dimer has 10 bound states. Finally, using the coupling potential strength, $V_{co}=3.883$ K, we get that $a_\mathrm{tw}=\infty$ for $d=3$.

We collect the details of this potential in the second row of Table~\ref{tab1},

\subsection{Dimension dependence of the Feshbach resonances}

Now that the parameters for the two-body interactions have been selected, we can investigate how the position of the 
Feshbach resonances change with dimension $d$.

\begin{figure}
\includegraphics[width=1\linewidth]{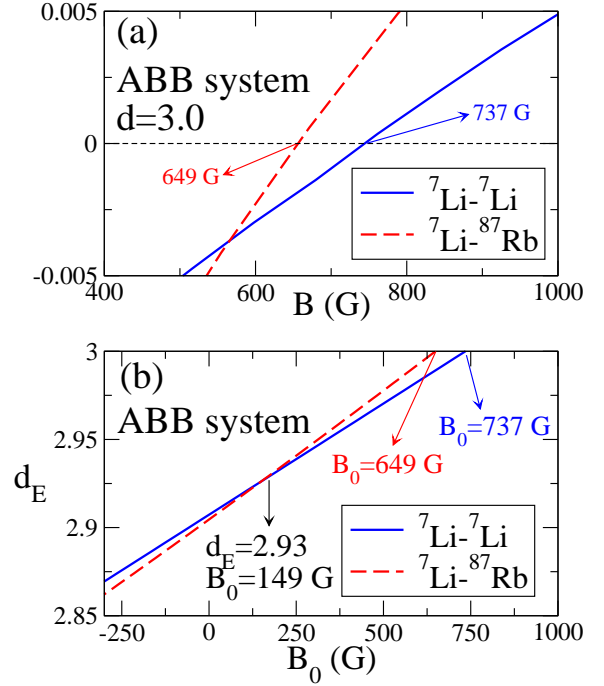}
\caption{(a) For $d=3$, the denominator of Eq.(\ref{eq25}) as a function
of the magnetic field $B$ for the $^7$Li-$^7$Li (solid) and $^7$Li-$^{87}$Rb (dashed) dimers. 
The crossings with the zero-axis correspond to the $B$-values 
producing a Feshbach resonance for $d=3$.
(b) $(B_0,d_E)$ values that give rise to a Feshbach resonance in the $^7$Li-$^7$Li (solid curves)
and $^7$Li-$^{87}$Rb (dashed curves) dimers. The point where the curves cross corresponds to the $(B_0,d_E)$
value producing a resonance simultaneously in both systems. }
\label{denom3}
\end{figure}

In Fig.~\ref{denom3}a we show, for $d=3$, the denominator of Eq.(\ref{eq25}), as a function of the magnetic field
$B$, for the potentials parameters for the $^{7}$Li-$^7$Li and $^{7}$Li-$^{87}$Rb 
dimers. The crossings of the curves with the zero-axis correspond to the $B$-values giving rise to $a_\mathrm{tw}=\infty$, 
and therefore producing a Feshbach resonance. These values of the magnetic field reproduce of course the
values given in Refs.~\cite{chi10} and \cite{mar09}, and quoted in Table~\ref{tab1}, 
$B_0=737$ G for $^7$Li-$^7$Li and $B_0=649$ G for $^7$Li-$^{87}$Rb.

The same calculation as in Fig.~\ref{denom3}a, but for different values of the dimension, would give rise to different
$B_0$-values such that a Feshbach resonance appears on each of the two-body systems. 
In Fig.~\ref{denom3}b we show the family $(B_0,d_E)$ values producing a Feshbach resonance for the $^7$Li-$^7$Li 
and the $^7$Li-$^{87}$Rb dimers, which are the ones entering in the $^{87}$Rb-$^7$Li-$^7$Li trimer. The arrows in the figure indicate the $B_0$ values 
obtained for $d_E=3$, $B_0=737$ G ($^7$Li-$^7$Li) and $649$ G ($^7$Li-$^{87}$Rb), which, as mentioned above,
are the ones used to construct the potentials.

The relevant thing is that, as we can see in the figure, the solid and dashed curves cross at some particular
$(B_0,d_E)$ point, $d_E=2.93$ and $B_0=149$ G. For these values of the dimension and the magnetic field, 
the scattering length, $a_\mathrm{tw}$, for both two-body potentials is, simultaneously, equal to infinity.
This fact implies that, for this specific
$(B_0,d_E)$ value, the three-body system $^{87}$Rb-$^7$Li-$^7$Li fulfills the Efimov conditions simultaneously in
all the internal two-body subsystems. In particular, for $d_E=2.93$ the energy scale factor for two consecutive
Efimov states is only about 317, a value that can not even be compared to the scale factor obtained
when only the  $^{87}$Rb-$^7$Li has infinite scattering length, since in this case the scale factor is larger than 
$10^{50}$. 

The price to pay is that, as discussed below Eq.(\ref{eq1}), a $d_E$ value of 2.93, which leads to $b_{ho}/r_0=3.71$,
amounts to confinement
frequencies in the range of the MHz, clearly higher than today's available experimental values.

\section{The ABC-system}
\label{sect7}

As shown in Fig.~\ref{fig0b}, for the case of three different constituents, the most interesting situation is when
one of them is clearly lighter than the other two. In this case, if the interaction between the light particle and
the other two is tuned to produce $a_\mathrm{tw}=\infty$ in both cases, then the Efimov conditions are fulfilled, and,
furthermore, the energy separation between the Efimov states is relatively small.

To illustrate this point we choose the $^7$Li-$^{39}$K-$^{87}$Rb system. The interaction used to describe the 
$^7$Li-$^{87}$Rb dimer was detailed in the previous section. The remaining interaction needed, the $^7$Li-$^{39}$K
potential, is described below.

\subsection{The $^7$Li-$^{39}$K dimer}

The details of the Feshbach resonance in three dimensions, required to determine the $^7$Li-$^{39}$K potential, are taken
from Ref.~\cite{cui18}. In particular the following information about the resonance is given: $B_0=319$ G, 
$\Delta B=30$ G, $a_\mathrm{open}=89 a_0$, and $\delta_\mu=1.5 \mu_B$. Also, from the value provided for the $C_6$ 
van der Waals coefficient we obtain the van der Waals length, $R_\mathrm{vdW}=42.1 a_0$, which is the value used for the  
radius of our square-well potentials. 

With these numbers we get $\delta_\mu B_0=0.032$ K, and, as in the previous cases, we choose $E^*=5$ K. This value
of $E^*$, together with the values given above, by means of Eq.(\ref{acl3d}), results in $a_\mathrm{closed}=1.71 a_0$.
 
As for the other two potentials, we adjust $V_o$ and $V_c$ such that they reproduce the required values of 
 $a_\mathrm{open}$ and $a_\mathrm{closed}$. We take then $V_o=17.072$ K and $V_c=17.040$ K. These two
strengths give rise to a $^7$Li-$^{39}$K dimer holding about 15 bound states. When combined with a 
 coupling strength $V_{co}=2.505$ K, we obtain a two-component scattering length $a_\mathrm{tw}=\infty$ for $d=3$.
 
 As for the other two potentials, we collect in the last row of Table~\ref{tab1} the details of this interaction.

\subsection{Dimension dependence of the Feshbach resonance positions}

\begin{figure}
\includegraphics[width=1\linewidth]{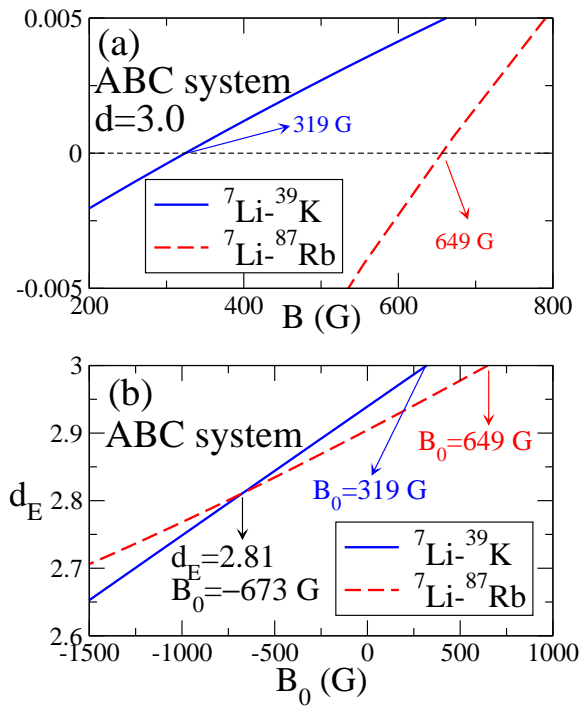}
\caption{The same as in Fig.~\ref{denom3} for the $^7$Li-$^{39}$K (solid) and $^7$Li-$^{87}$Rb (dashed) dimers.}
\label{fig1}
\end{figure}

We proceed now as for the ABB-case, and show in Fig.~\ref{fig1}a the denominator of Eq.(\ref{eq25}), as a 
function of the magnetic field, $B$, for the potential described above for the $^7$Li-$^{39}$K dimer and $d=3$. 
We also show this denominator for the $^7$Li-$^{87}$Rb case, already plotted in
Fig.~\ref{denom3}a . Again, the crossing of each curve with the zero-axis corresponds to the 
different $B$-values giving rise to $a_\mathrm{tw}=\infty$, and therefore producing a Feshbach resonance. 
This crossings can be easily seen in the zoom of the figure shown in the inset. Again, as expected, the crossings
correspond to the values $B_0=319$ G ($^7$Li-$^{39}$K) and $B_0=649$ G ($^7$Li-$^{87}$Rb) used to construct
the potentials (see Table~\ref{tab1}).

We can now investigate how the position of the Feshbach resonances change with the dimension $d$, and
get the different $(B_0,d_E)$ values that produce a Feshbach resonance in both, the $^7$Li-$^{39}$K
and the $^{7}$Li-$^{87}$Rb dimers, and in particular, check if it is possible to find a $(B_0,d_E)$ value
that produces a resonance simultaneously in both systems. 

This is shown in Fig.~\ref{fig1}b, which is the same as in Fig.~\ref{denom3}b, but for the $^7$Li-$^{39}$K 
and $^{7}$Li-$^{87}$Rb dimers. We can see that, again, both curves cross, determining therefore
a dimension, $d_E=2.81$, and a magnetic field, $B_0=-673$ G, such that $a_\mathrm{tw}=\infty$ in the two systems 
at the same time. The negative value of $B_0$ means a projection of the magnetic field opposite to the direction of
the magnetic moment difference of the open and closed two-body channels.
The Efimov conditions are therefore fulfilled, and with these mass ratios for $d=2.81$, and assuming that
the scattering length between the $^{39}$K and $^{87}$Rb is small, the energy scale factor between two
consecutive Efimov states is then about 330. From Eq.(\ref{eq1}) we obtain that this dimension corresponds to 
$b_{ho}/r_0=2.17$, which again represents a pretty tight confinement, with frequencies in the range of the MHz.

\section{Summary and conclusions}           
\label{sect8} 

The $d$-method, where the dimension $d$ can take non-integer values, is an efficient way of describing systems
squeezed by an external field. During the squeezing process, the effective interaction between two particles
is modified, in such a way that, if the two-body system is unbound in three dimensions, there must necessarily be
a moment where the two-body binding energy is equal to zero (infinite scattering length). The equivalent dimension
describing this squeezing situation is what we call the critical dimension, $d_E$.

A different procedure that permits as well the modification of the effective two-body interaction consists on 
putting the system under the effect of a static magnetic field. The coupling of the field with the magnetic
moments of the particles allows in particular the population of the so called Feshbach resonances, which
appear for some particular value, $B_0$, of the magnetic field, and which corresponds as well to an infinite two-body scattering length.

In this work we have considered the combination of these two different procedures in order to investigate the
possibility of simultaneous tuning of two of the different interactions involved in an
asymmetric three-body system. This would permit to construct ABB three-body systems where the AB and BB two-body
potentials have both infinite scattering length. In this way, for systems consisting of two identical light particles and 
a heavy one, the energy scale factor between consecutive Efimov states is reduced by many orders of magnitude.
Even for systems made of three different particles, ABC systems, it would be possible to tune the AB and AC potentials,
and construct in this way a series of Efimov states whose energy separation is reasonably small when the mass
of particle A is much smaller than the mass of the other two. 

The analysis of the Feshbach resonances as a function of the dimension has been done analytically assuming 
a two-component problem for the two-body systems, where only one open and one closed channel contribute,
and assuming square-well two-body potentials. In this way we have derived analytic expressions for the open-channel
and closed-channel wave functions, and from them we have obtained an analytic expression for the phase shifts.
By expansion of the phase shift in powers of energy we have obtained analytic expressions for the scattering
length and the effective range. A detailed analysis of the two-component scattering length allowed us to extract
the curve made by all the $(B_0,d_E)$ points such that for that value, $B_0$, of the magnetic field, and that
dimension, $d_E$, the two-body scattering length is equal to infinity. 

All these analytical expressions have been used to investigate the two-body potentials involved in two different
three-body systems, $^{87}$Rb-$^7$Li-$^7$Li, as an example of an ABB system containing two light particles and a
heavy one, and $^7$Li-$^{39}$K-$^{87}$Rb, as an example of an ABC system where one particle is much lighter than the
other two. Making use of the available experimental information we have constructed the $^{87}$Rb-$^7$Li, 
$^7$Li-$^7$Li, and $^7$Li-$^{39}$K potentials in three dimensions. With the obtained potential parameters
we have then made use of the derived analytical expression for the scattering length and, for each of the
interactions, we have constructed the $(B_0,d_E)$-curve providing all the $(B_0,d_E)$ pair values
corresponding to infinite two-body scattering length. We have then observed that the curves arising from the
$^{87}$Rb-$^7$Li and $^7$Li-$^7$Li potential cross at some particular $(B_0,d_E)$-point. For these values
of the magnetic field and the dimension, both the $^{87}$Rb-$^7$Li and the $^7$Li-$^7$Li, scattering lengths
are equal to infinity, and the energy scale factor between two consecutive Efimov states reduce by many orders 
of magnitude down to a value of about 317. The same happens for the  $^7$Li-$^{39}$K-$^{87}$Rb system.
The $(B_0,d_E)$-curves corresponding to the $^{87}$Rb-$^7$Li and the $^7$Li-$^{39}$K potentials cross as well
at some particular $(B_0,d_E)$-point, in such a way that for this magnetic
field and dimension the Efimov conditions are fulfilled, and the separation between the Efimov states reduces to a factor of
about 330.

The trimers chosen in this work and the rough estimate of the atom-atom potentials, have led to confinement
frequencies that are too high compared to the current available experimental values. However, this fact does
not invalidate the main conclusion, which is that simultaneous tuning of two different two-body potentials after 
combination of squeezing of the system and the population of Feshbach resonances by means of a static magnetic 
field is in principle possible. This could be today limited to systems where this happens for a very modest 
confinement strength, but there is always the possibility that, eventually, much higher frequencies could be reached.

Our results open the door to the construction of Efimov states in three-body systems which
in principle are not good candidates to exhibit this phenomenon, like systems made of two identical particles
and a heavy one, or even systems made of three different particles.

In summary, the numerical discussions are based on short-range
particle-particle square-well potentials.  However, the Efimov physics
is a long-distance phenomenon, and appearance and properties are
independent of the employed short-range potentials. We emphasize
therefore that our conclusions are universal as described in terms of
the long-range gross properties, scattering length and effective
range.  The results are also universal in the sense that they apply to
all two and three-body systems in any subfield of physics or
chemistry.

A number of perspectives are immediate for future few-body
investigations involving external deformed squeezing potentials, which
can be studied using the equivalent $d$-method for non-integer
dimensions.  More precise calculations on specific experimental
selected systems could be necessary.  Also extended computations would
be of interest, for example shapes of Feshbach resonances and
three-body recombination rates related to Efimov physics.

\bmhead{Acknowledgments}
This work has been partially supported by grant No.
PGC2018-093636-B-I00, funded by MCIN/AEI/10.13039/501100011033 and
by ``ERDF A way of making Europe''.
We acknowledge Jan Arlt for providing experimental information.

\section*{Author contributions}

Both authors have equally contributed to the discussion and the 
analytic calculations. EG has performed the numerical calculations.

\vspace*{5mm}
\noindent
{\bf Data Availability Statement} This manuscript has no
associated data.

\begin{appendices}

\section{Matching at box radius, $x=1$.}
\label{appmatch}

The logarithmic derivatives of the solutions in Eqs.(\ref{e110}) and
(\ref{e250}), and in Eqs.(\ref{e115}) and (\ref{e255}), outside and
inside the box, must be equal at the box radius, $x=1$.  This condition
leads to the following two equations:
\begin{eqnarray} \label{e410}
  && i\kappa_{c} \frac{h_\ell^{(+)\prime}(i\kappa_{c})}{h_\ell^{(+)}(i\kappa_{c})} =
  \\ \nonumber  &&
  \frac{(D^2_+-a_-) C_+D_+ j_\ell^{\prime}(D_+) + a_{co} D_-C_- j_\ell^{\prime}(D_-)}
    { (D^2_+-a_-) C_+ j_\ell(D_+) + a_{co} C_- j_\ell(D_-)}  \;,
\end{eqnarray}
\begin{eqnarray} \label{e420}
  && k_o \frac{\cos\delta j_\ell^{\prime}(k_{o}) - \sin\delta \eta_\ell^{\prime}(k_{o})}
  {\cos\delta j_\ell(k_{o}) -\sin\delta \eta_\ell(k_{o})  } =
  \\ \nonumber &&
  \frac{(D^2_--a_+) C_-D_- j_\ell^{\prime}(D_-) + a_{co} D_+C_+ j_\ell^{\prime}(D_+)} 
 { (D^2_--a_+) C_- j_\ell(D_-) + a_{co} C_+ j_\ell(D_+)} \; ,
\end{eqnarray}
where the prime denotes derivative with respect to the full coordinate of
the corresponding Hankel or Bessel function. Both equations contain the ratio 
between the normalization constants, $C_+$ and $C_-$ in Eq.(\ref{e130}).
Eq.(\ref{e410}) can be used to determine
\begin{equation} \label{e430}
  \frac{C_+}{C_-}  = \frac{a_{co}}{(a_- - D^2_+)} \frac{j_\ell(D_-)}{j_\ell(D_+) }
 \frac{i\kappa_{c} \frac{h_\ell^{(+)\prime}(i\kappa_{c})}{h_\ell^{(+)}(i\kappa_{c})}
   -D_- \frac{j_\ell^{\prime}(D_-)}{j_\ell(D_-)}} 
   {i\kappa_{c} \frac{h_\ell^{(+)\prime}(i\kappa_{c})}{h_\ell^{(+)}(i\kappa_{c})}
     - D_+ \frac{j_\ell^{\prime}(D_+)}{j_\ell(D_+)}} \; ,
\end{equation}
which by use of Eq.(\ref{A480}) reduces to
\begin{eqnarray} \label{e433}
 \frac{C_+}{C_-}  = F_1 \frac{a_{co}}{(a_- - D^2_+)} \frac{j_\ell(D_-)}{j_\ell(D_+) },
\end{eqnarray}
where we have defined
\begin{eqnarray} \nonumber
  F_1 &=&\frac{D_- \frac{j_{\ell+1}(D_-)}{j_\ell(D_-)}
    - i\kappa_{c} \frac{h_{\ell+1}^{(+)}(i\kappa_{c})}{h_\ell^{(+)}(i\kappa_{c})} }
       { D_+ \frac{j_{\ell+1}(D_+)}{j_\ell(D_+)} 
    -i \kappa_{c} \frac{h_{\ell+1}^{(+)}(i\kappa_{c})}{h_\ell^{(+)}(i\kappa_{c})}} \\
  &=& \frac{D_- \frac{j_{\ell+1}(D_-)}{j_\ell(D_-)}
  - \kappa_{c} \frac{K_{\ell+\frac{3}{2}}(\kappa_{c})}{K_{\ell+\frac{1}{2}}(\kappa_{c})}}
  { D_+ \frac{j_{\ell+1}(D_+)}{j_\ell(D_+)} 
 - \kappa_{c} \frac{K_{\ell+\frac{3}{2}}(\kappa_{c})}{K_{\ell+\frac{1}{2}}(\kappa_{c})}},
 \label{e436}
\end{eqnarray}
and where the relation in Eq.(\ref{A531}) has been used.

From Eq.(\ref{e420}) we can now extract the phase shift and find
\begin{eqnarray} \label{e440}
  \cot\delta =  \frac{\eta_\ell(k_{o})}{j_\ell(k_{o})}
  \frac{F_2 - k_{o} \frac{\eta_\ell^{\prime}(k_{o})}{\eta_\ell(k_{o})}}
  {F_2 - k_{o} \frac{j_\ell^{\prime}(k_{o})}{j_\ell(k_{o})}} \; ,
\end{eqnarray}
where
\begin{equation} \label{e450}
  F_2 = \frac{ (D^2_--a_+) D_- j_\ell^{\prime}(D_-)  + \frac{C_+}{C_-} a_{co} D_+ j_\ell^{\prime}(D_+)}
  { (D^2_--a_+) j_\ell(D_-)  + \frac{C_+}{C_-} a_{co} j_\ell(D_+)} \; .
\end{equation}

Using Eq.(\ref{e433}) and, again Eq.(\ref{A480}), we reformulate Eqs. (\ref{e440}) and
(\ref{e450}), and finally write:
\begin{eqnarray} \label{e443ap}
  \cot\delta =  \frac{\eta_\ell(k_{o})}{j_\ell(k_{o})}
  \frac{F_2  - k_{o} \frac{\eta_{\ell+1}(k_{o})}{\eta_\ell(k_{o})}}
  {F_2 - k_{o} \frac{j_{\ell+1}(k_{o})}{j_\ell(k_{o})}} \; ,
\end{eqnarray}
where, after inserting the expression for $C_+/C_-$ from Eq.(\ref{e433}),
and using Eqs.(\ref{e152}) and (\ref{A480}), we have redefined $F_2$ as:
\begin{equation} \label{e456}
 F_2  =
   \frac{ D_- \frac{j_{\ell+1}(D_-)}{j_{\ell}(D_-)}
    + F_1 \frac{a_{co}^2}{(D^2_--a_+)^2} D_+ \frac{j_{\ell+1}(D_+)}{j_{\ell}(D_+)}}
  {1 + F_1 \frac{a_{co}^2}{(D^2_--a_+)^2}  }  \;,
\end{equation}
which, together with Eq.(\ref{e436}), provides 
the final analytical formula for the phase shift. 

Note that in case of no coupling, $a_{co}=0$, we have that 
$F_2=\sqrt{a_-} j_{\ell+1}(\sqrt{a_-})/j_{\ell}(\sqrt{a_-})$, 
and Eq.(\ref{e443ap}) reduces to Eq.(20) of Ref.~\cite{chr22}.

\section{Low-energy expansion of the phase shift}
\label{apareff}

The quantities entering in the expression for the phase shift, 
Eq.(\ref{e443ap}), are
$D_{\pm}$ and $a_{\pm}$ in various combinations. The relevant
expressions are Eqs.(\ref{e87}), (\ref{e140}), (\ref{e141}), (\ref{e320}), 
and (\ref{e322}). The following abbreviations
in terms of the basic interaction quantities are useful
\begin{eqnarray} 
& &  b_0 =  \sqrt{(a_+ - a_-)^2 + 4 a_{co}^2}  \;,  \\
& &  b_1 = S^2_c+S^2_o -\frac{2\mu R_0^2 E^*}{\hbar^2} = (a_+ + a_-)^{(E=0)}\;,  \\
& & D_{\pm}^{(E=0)} \equiv  D_\pm^{(0)} = \sqrt{(b_1 \pm b_0)/2} \;,   \label{e371} \\
& & \kappa_c^{(E=0)}\equiv \kappa_c^{(0)} = \sqrt{\frac{2\mu R_0^2 E^*}{\hbar^2}} \;, \label{e372} \\
& & a_+^{(E=0)}\equiv a_+^{(0)}=S_c^2-\kappa_c^{(0)2}. \label{e373}
\end{eqnarray}

The expressions in Eq.(\ref{e443ap}) can be reduced for low energy as \cite{abr65}
\begin{equation}
  \frac{\eta_{\ell}(k_{o})}{j_\ell(k_{o})} \rightarrow
 -  \frac{\left[(2\ell+1)!!\right]^2}{(2\ell+1) k_o^{2\ell+1}}
  \bigg(1 + \frac{(2\ell +1) k^2_{o}}{(2\ell -1)(2\ell +3)}\bigg),
\end{equation}
\begin{equation}
  k_{o} \frac{\eta_{\ell+1}(k_{o})}{\eta_\ell(k_{o})} \rightarrow (2\ell+1)
  \left(1- \frac{k^2_{o}}{(2\ell -1)(2\ell +1)}\right) \; ,
 \label{e377}
\end{equation}
and
\begin{equation}
 k_{o} \frac{j_{\ell+1}(k_{o})}{j_\ell(k_{o})} \rightarrow
   \frac{k_{o}^2}{2\ell+3}. 
\end{equation}

The remaining functions to expand around finite values are then $F_1$,
$F_2$, and the ratios of various Bessel functions.  With obvious
notation we can write
\begin{equation} \label{e831}
 F_1 \rightarrow F_1^{(E=0)} + E  \left. {\frac{\partial F_1}{\partial E}}\right\vert_{E=0} 
 \equiv F_1^{(0)} + E F_1^{\prime(0)} \; ,  
 \end{equation}
 \begin{equation} \label{e832}
 F_2 \rightarrow F_2^{(E=0)} + E \left. {\frac{\partial F_2}{\partial E}}\right\vert_{E=0}
 \equiv F_2^{(0)} + E F_2^{\prime (0)} \; ,
\end{equation}
where we explicitly have for both Eqs.(\ref{e436}) and (\ref{e456}) 
\begin{equation} \label{e537}
  F_1^{(E=0)}\equiv F_1^{(0)} = \frac{D_-^{(0)} \frac{j_{\ell+1}(D_{-}^{(0)})}{j_\ell(D_-^{(0)})}
    -\kappa_c^{(0)} \frac{K_{\ell+3/2}^{(+)}(\kappa_c^{(0)})}{K_{\ell+1/2}(\kappa_c^{(0)})} }
 {D_+^{(0)} \frac{j_{\ell+1}(D_+^{(0)})}{j_\ell(D_+^{(0)})} 
     -\kappa_c^{(0)} \frac{K_{\ell+3/2}^{(+)}(\kappa_c^{(0)})}{K_{\ell+1/2}^{(+)}(\kappa_c^{(0)})}} \; .
\end{equation}
and
\begin{eqnarray} \label{e577}
\lefteqn{  F_2^{(E=0)} \equiv F_2^{(0)} = } \\ & &
   \frac{ D_-^{(0)} \frac{j_{\ell+1}(D_-^{(0)})}{j_{\ell}(D_-^{(0)})}
   + F_1^{(0)} \frac{a_{co}^2}{(D_-^{(0)2}-a_+^{(0)})^2} D_+^{(0)}
   \frac{j_{\ell+1}(D_+^{(0)})}{j_{\ell}(D_+^{(0)})}}
  {1 + F_1^{(0)} \frac{a_{co}^2}{(D^{(0)2}_--a_+^{(0)})^2}  } \nonumber \; .
\end{eqnarray}

The derivatives, $F_1^{\prime(0)}$ and $F_2^{\prime(0)}$ both involve
combinations of derivatives of ratios of spherical Bessel or Hankel functions of the form
$f_a(z) = z \frac{j_{a+1}(z)}{j_{a}(z)}$, where $z$ is $D_{\pm}$, and 
$g_a(z) = z \frac{h_{a+1}^{(+)}(z)}{h_{a}^{(+)}(z)}$ with $z=i \kappa_c$.  The
derivatives, $f^\prime_a(z)$ and $g^\prime_a(z)$, of these pieces are evaluated
by use of Eq.(\ref{A543}), where the derivatives of $z$ must also be
inserted, that is
\begin{eqnarray} \label{e587}
 \frac{\partial D_{\pm}}{\partial E} = \frac{\mu  R_0^2}{D_{\pm} \hbar^2}\;,\; 
 \frac{\partial (i \kappa_c)}{\partial E} =  \frac{\mu  R_0^2}{i\kappa_c\hbar^2}\;.
\end{eqnarray}

Noting also that $(D^2_--a_+)$ is $E$-independent, the derivatives are then easily obtained to be
\begin{eqnarray}  \label{e604}
  && \frac{F_2^{\prime(0)}}{F_2^{(0)}} =
  -  \frac{F_1^{\prime(0)}a_{co}^2/(D^{(0)2}_--a_+^{(0)})^2}{1+ F_1^{(0) } a_{co}^2/(D^{(0)2}_--a_+^{(0)})^2} \\ \nonumber 
  &+& \frac{f^\prime_{\ell}(D^{(0)}_-) +  \frac{a_{co}^2} {(D^{(0)2}_--a_+^{(0)})^2}
    \big(F_1^{(0)} f^\prime_{\ell}(D^{(0)}_+) +  F_1^{\prime(0)} f_{\ell}(D^{(0)}_+)\big)}
 {f_{\ell}(D^{(0)}_-) + F_1^{(0)} f_{\ell}(D^{(0)}_+)
   \frac{a_{co}^2}{(D^{(0)2}_--a_+^{(0)})^2}} \;,  
\end{eqnarray}
and
\begin{eqnarray} 
  \frac{F_1^{\prime(0)}}{F_1^{(0)}} &=& \frac{f^\prime_{\ell}(D^{(0)}_-)
  - g^\prime_{\ell}(i\kappa^{(0)}_c)} {f_{\ell}(D^{(0)}_-) - g_{\ell}(i\kappa^{(0)}_c)}
    \nonumber \\ \label{e607} &-& \frac{f^\prime_{\ell}(D^{(0)}_+)
 - g^\prime_{\ell}(i\kappa^{(0)}_c)} {f_{\ell}(D^{(0)}_+) - g_{\ell}(i\kappa^{(0)}_c)}\;.
\end{eqnarray}
  
The phase shift expanded to second order in energy is now
\begin{eqnarray} 
 && \frac{(2\ell+1)}{[(2\ell+1)!!]^2} k_o^{2\ell+1}  \cot\delta \simeq
  -\bigg(1+ \frac{(2\ell +1) k_o^2}{(2\ell-1) (2\ell+3)}  \bigg)  \nonumber \\
  &&\times \bigg(\frac{F_2^{(0)} -(2\ell+1) + E (F_2^{\prime(0)} +
   \frac{k_o^2}{E} \frac{1}{2\ell -1})}
 {F_2^{(0)} +E (F_2^{\prime(0)}-\frac{k_o^2}{E}\frac{1}{2\ell+3})}\bigg),
   \label{e597}
\end{eqnarray}
which can be rewritten as
\begin{eqnarray} \label{e666}
 && \frac{(2\ell+1)}{[(2\ell+1)!!]^2} k_o^{2\ell+1}  \cot\delta \simeq
  -\bigg(1 - \frac{2\ell +1}{F_2^{(0)}}\bigg)
  \\ \nonumber &&
   - E \,\, \frac{2\ell+1}{F_2^{(0)}}\bigg[
  \frac{F_2^{\prime(0)}}{F_2^{(0)}} - \frac{k_o^2}{E} \frac{1}{2\ell+3}
  \left(1+\frac{1}{F_2^{(0)}}-\frac{F_2^{(0)}}{2\ell-1} \right) 
  \bigg],
\end{eqnarray}
where for non-integer values of $\ell$ the factor $(2\ell+1)!!$
should be replaced by $\Gamma(2\ell+2)/(2^\ell\Gamma(\ell+1))$.

\section{The two-component effective range}
\label{appc}

As done for the scattering length, after comparison of Eqs.(\ref{delexp}) and Eq.(\ref{e667}),
we obtain:
\begin{eqnarray}   \label{e676}
\lefteqn{  \frac{R_\mathrm{eff}}{R_0}  =
\frac{\hbar^2}{\mu R_0^2} \left( \frac{a_\mathrm{tw}}{R_0} \right)^{2\ell}   }
\\ & & 
\times  \frac{1}{F_2^{(0)}} \bigg[
  \frac{k_o^2}{E} \frac{1}{2\ell+3}
  \left(1+\frac{1}{F_2^{(0)}}-\frac{F_2^{(0)}}{2\ell-1} \right) - \frac{F_2^{\prime(0)}}{F_2^{(0)}}
  \bigg]. \nonumber
\end{eqnarray}

If we assume relative $s$-waves ($\ell=0$), making use of Eqs.(\ref{e604}) and (\ref{A480}), and after some tedious algebra, we can see that Eq.(\ref{e676}) reduces 
to the known analytic form of the effective range for the square-well potential \cite{dui04}:
\begin{equation}
\frac{R_ {\mathrm{eff}}}{R_0}  = 1+\frac{ 3 \tan S_o -3S_o - S_o^3 }{3 S_o (S_o - \tan S_o )^2 }.
\end{equation}

Making use of Eqs.(\ref{e577}), (\ref{acl}), (\ref{aop}), and (\ref{eq47}), we can obtain the 
effective range, Eq.(\ref{e676}), in terms of $a_\mathrm{open}$ and $a_\mathrm{closed}$,
although the expression obtained is certainly more complicated than the one given in 
Eq.(\ref{eq93}) for the scattering length.
  
\section{Some useful Bessel function properties}

Here, we collect relations between the Bessel functions
used to derive the expressions obtained in this work. They have
been taken from \cite{abr65}.

Below, $B_\ell$ represents either the regular, $j_\ell$, the
irregular, $\eta_\ell$, spherical Bessel functions, or the Hankel,
$h^{(\pm)}_{\ell}= j_\ell \pm i \eta_\ell$, functions.  The derivative
of $B_\ell(z)$ is given by:
\begin{eqnarray}     \label{A480}
  z \frac{d B_\ell(z)}{d z} = \ell B_\ell(z) - z B_{\ell+1}(z).  \label{A490}
\end{eqnarray}

The Hankel function $h_\ell^{(+)}(i\kappa_c)$ is related to the modified Bessel function
of third kind, $K_{\ell+\frac{1}{2}}(\kappa_c)$, by:
\begin{equation} \label{A501}
h_\ell^{(+)}(i\kappa_c) \propto \frac{1}{\sqrt{\kappa_c}} e^{-i\frac{\pi}{2}\ell} K_{\ell+\frac{1}{2}}(\kappa_c),
\end{equation}
which permits to write:
\begin{equation}\label{A531}
i\kappa_{c} \frac{h_{\ell+1}^{(+)}(i\kappa_{c})}{h_{\ell}^{(+)}(i\kappa_{c})}=
\kappa_{c} \frac{K_{\ell+\frac{3}{2}}(\kappa_{c})}{K_{\ell+\frac{1}{2}}(\kappa_{c})}.
\end{equation}

The derivative of $f_a(z) = z \frac{B_{a+1}}{B_{a}}$ is given by:
\begin{equation}     \label{A543}
  \frac{\partial f_a(z)}{\partial z} =
  z + z \bigg( \frac{B_{a+1}}{B_{a}}\bigg)^2 - (2 a +1)\frac{B_{a+1}}{B_{a}}.
\end{equation}

\end{appendices}


\bibliography{bibfile}


\begin{thebibliography}{22}
\ifx \bisbn   \undefined \def \bisbn  #1{ISBN #1}\fi
\ifx \binits  \undefined \def \binits#1{#1}\fi
\ifx \bauthor  \undefined \def \bauthor#1{#1}\fi
\ifx \batitle  \undefined \def \batitle#1{#1}\fi
\ifx \bjtitle  \undefined \def \bjtitle#1{#1}\fi
\ifx \bvolume  \undefined \def \bvolume#1{\textbf{#1}}\fi
\ifx \byear  \undefined \def \byear#1{#1}\fi
\ifx \bissue  \undefined \def \bissue#1{#1}\fi
\ifx \bfpage  \undefined \def \bfpage#1{#1}\fi
\ifx \blpage  \undefined \def \blpage #1{#1}\fi
\ifx \burl  \undefined \def \burl#1{\textsf{#1}}\fi
\ifx \doiurl  \undefined \def \doiurl#1{\url{https://doi.org/#1}}\fi
\ifx \betal  \undefined \def \betal{\textit{et al.}}\fi
\ifx \binstitute  \undefined \def \binstitute#1{#1}\fi
\ifx \binstitutionaled  \undefined \def \binstitutionaled#1{#1}\fi
\ifx \bctitle  \undefined \def \bctitle#1{#1}\fi
\ifx \beditor  \undefined \def \beditor#1{#1}\fi
\ifx \bpublisher  \undefined \def \bpublisher#1{#1}\fi
\ifx \bbtitle  \undefined \def \bbtitle#1{#1}\fi
\ifx \bedition  \undefined \def \bedition#1{#1}\fi
\ifx \bseriesno  \undefined \def \bseriesno#1{#1}\fi
\ifx \blocation  \undefined \def \blocation#1{#1}\fi
\ifx \bsertitle  \undefined \def \bsertitle#1{#1}\fi
\ifx \bsnm \undefined \def \bsnm#1{#1}\fi
\ifx \bsuffix \undefined \def \bsuffix#1{#1}\fi
\ifx \bparticle \undefined \def \bparticle#1{#1}\fi
\ifx \barticle \undefined \def \barticle#1{#1}\fi
\bibcommenthead
\ifx \bconfdate \undefined \def \bconfdate #1{#1}\fi
\ifx \botherref \undefined \def \botherref #1{#1}\fi
\ifx \url \undefined \def \url#1{\textsf{#1}}\fi
\ifx \bchapter \undefined \def \bchapter#1{#1}\fi
\ifx \bbook \undefined \def \bbook#1{#1}\fi
\ifx \bcomment \undefined \def \bcomment#1{#1}\fi
\ifx \oauthor \undefined \def \oauthor#1{#1}\fi
\ifx \citeauthoryear \undefined \def \citeauthoryear#1{#1}\fi
\ifx \endbibitem  \undefined \def \endbibitem {}\fi
\ifx \bconflocation  \undefined \def \bconflocation#1{#1}\fi
\ifx \arxivurl  \undefined \def \arxivurl#1{\textsf{#1}}\fi
\csname PreBibitemsHook\endcsname

\bibitem{ino98}
\begin{barticle}
\bauthor{\bsnm{Inouye}, \binits{S.}},
\bauthor{\bsnm{Andrews}, \binits{M.R.}},
\bauthor{\bsnm{Stenger}, \binits{J.}},
\bauthor{\bsnm{Miesner}, \binits{H.-J.}},
\bauthor{\bsnm{Stamper-Kurn}, \binits{D.M.}},
\bauthor{\bsnm{Ketterle}, \binits{W.}}:
\batitle{{Observation of Feshbach resonances in a Bose-Einstein condensate}}.
\bjtitle{Nature}
\bvolume{392},
\bfpage{151}
(\byear{1998}).
\doiurl{10.1038/32354}
\end{barticle}
\endbibitem

\bibitem{chi10}
\begin{barticle}
\bauthor{\bsnm{Chin}, \binits{C.}},
\bauthor{\bsnm{Grimm}, \binits{R.}},
\bauthor{\bsnm{Julienne}, \binits{P.}},
\bauthor{\bsnm{Tiesinga}, \binits{E.}}:
\batitle{{Feshbach resonances in ultracold gases}}.
\bjtitle{Rev. Mod. Phys.}
\bvolume{82},
\bfpage{1225}
(\byear{2010}).
\doiurl{10.1103/RevModPhys.82.1225}
\end{barticle}
\endbibitem

\bibitem{efi70}
\begin{barticle}
\bauthor{\bsnm{Efimov}, \binits{V.}}:
\batitle{{Energy levels arising from resonant two-body forces in a three-body
  system}}.
\bjtitle{Phys. Lett. B}
\bvolume{33},
\bfpage{563}
(\byear{1970}).
\doiurl{10.1016/0370-2693(70)90349-7}
\end{barticle}
\endbibitem

\bibitem{hua14}
\begin{barticle}
\bauthor{\bsnm{Huang}, \binits{B.}},
\bauthor{\bsnm{Sidorenkov}, \binits{L.A.}},
\bauthor{\bsnm{Grimm}, \binits{R.}},
\bauthor{\bsnm{Hutson}, \binits{J.M.}}:
\batitle{{Observation of the second triatomic resonance in Efimov's scenario}}.
\bjtitle{Phys. Rev. Lett.}
\bvolume{112},
\bfpage{190401}
(\byear{2014}).
\doiurl{10.1103/PhysRevLett.112.190401}
\end{barticle}
\endbibitem

\bibitem{tun14}
\begin{barticle}
\bauthor{\bsnm{Tung}, \binits{S.-K.}},
\bauthor{\bsnm{Jim{\'{e}}nez-Garc{\'{i}}a}, \binits{K.}},
\bauthor{\bsnm{Johansen}, \binits{J.}},
\bauthor{\bsnm{Parker}, \binits{C.V.}},
\bauthor{\bsnm{Chin}, \binits{C.}}:
\batitle{{Geometric scaling of Efimov states in a $^6$Li-$^{133}$Cs mixture}}.
\bjtitle{Phys. Rev. Lett.}
\bvolume{113},
\bfpage{240402}
(\byear{2014}).
\doiurl{10.1103/PhysRevLett.113.240402}
\end{barticle}
\endbibitem

\bibitem{gar21}
\begin{barticle}
\bauthor{\bsnm{Garrido}, \binits{E.}},
\bauthor{\bsnm{Jensen}, \binits{A.S.}}:
\batitle{{Efimov effect in non-integer dimensions induced by an external
  field}}.
\bjtitle{Phys. Lett. A}
\bvolume{385},
\bfpage{126982}
(\byear{2021}).
\doiurl{10.1016/j.physleta.2020.126982}
\end{barticle}
\endbibitem

\bibitem{gar19a}
\begin{barticle}
\bauthor{\bsnm{Garrido}, \binits{E.}},
\bauthor{\bsnm{Jensen}, \binits{A.S.}},
\bauthor{\bsnm{{\'{A}}lvarez-Rodr{\'{i}}guez}, \binits{R.}}:
\batitle{{Few-body quantum method in a $d$-dimensional space}}.
\bjtitle{Phys. Lett. A}
\bvolume{383},
\bfpage{2021}
(\byear{2019}).
\doiurl{10.1016/j.physleta.2019.04.007}
\end{barticle}
\endbibitem

\bibitem{nis09}
\begin{barticle}
\bauthor{\bsnm{Nishida}, \binits{Y.}},
\bauthor{\bsnm{Tan}, \binits{S.}}:
\batitle{{Confinement-induced Efimov resonances in Fermi-Fermi mixtures}}.
\bjtitle{Phys. Rev. A}
\bvolume{79},
\bfpage{060701}
(\byear{2009}).
\doiurl{10.1103/PhysRevA.79.060701}
\end{barticle}
\endbibitem

\bibitem{gar19b}
\begin{barticle}
\bauthor{\bsnm{Garrido}, \binits{E.}},
\bauthor{\bsnm{Jensen}, \binits{A.S.}}:
\batitle{{Confinement of two-body systems and calculations in $d$ dimensions}}.
\bjtitle{Phys. Rev. Research}
\bvolume{1},
\bfpage{23009}
(\byear{2019}).
\doiurl{10.1103/PhysRevResearch.1.023009}
\end{barticle}
\endbibitem

\bibitem{gar20}
\begin{barticle}
\bauthor{\bsnm{Garrido}, \binits{E.}},
\bauthor{\bsnm{Jensen}, \binits{A.S.}}:
\batitle{{Three identical bosons : Properties in noninteger dimensions and in
  external fields}}.
\bjtitle{Phys. Rev. Research}
\bvolume{2},
\bfpage{033261}
(\byear{2020}).
\doiurl{10.1103/PhysRevResearch.2.033261}
\end{barticle}
\endbibitem

\bibitem{gar21b}
\begin{barticle}
\bauthor{\bsnm{Garrido}, \binits{E.}},
\bauthor{\bsnm{Jensen}, \binits{A.S.}}:
\batitle{{Efimov effect evaporation after confinemen}}.
\bjtitle{Few-body Syst.}
\bvolume{62},
\bfpage{25}
(\byear{2021}).
\doiurl{10.1007/s00601-021-01613-4}
\end{barticle}
\endbibitem

\bibitem{lan09}
\begin{barticle}
\bauthor{\bsnm{Lange}, \binits{A.D.}},
\bauthor{\bsnm{Pilch}, \binits{K.}},
\bauthor{\bsnm{Prantner}, \binits{A.}},
\bauthor{\bsnm{Ferlaino}, \binits{F.}},
\bauthor{\bsnm{Engeser}, \binits{B.}},
\bauthor{\bsnm{N\"{a}gerl}, \binits{H.-C.}},
\bauthor{\bsnm{Grimm}, \binits{R.}},
\bauthor{\bsnm{Chin}, \binits{C.}}:
\batitle{{Determination of atomic scattering lengths from measurements of
  molecular binding energies near Feshbach resonances}}.
\bjtitle{Phys. Rev. A}
\bvolume{79},
\bfpage{013622}
(\byear{2009}).
\doiurl{10.1103/PhysRevA.79.013622}
\end{barticle}
\endbibitem

\bibitem{soe13}
\begin{barticle}
\bauthor{\bsnm{S{\o}rensen}, \binits{P.K.}},
\bauthor{\bsnm{Fedorov}, \binits{D.V.}},
\bauthor{\bsnm{Jensen}, \binits{A.S.}}:
\batitle{{Three-body recombination rates near a Feshbach resonance within a
  two-channel contact interaction model}}.
\bjtitle{Few-body Syst.}
\bvolume{54},
\bfpage{579}
(\byear{2013}).
\doiurl{10.1007/s00601-012-0312-7}
\end{barticle}
\endbibitem

\bibitem{mar04}
\begin{barticle}
\bauthor{\bsnm{Marcelis}, \binits{B.}},
\bauthor{\bparticle{van} \bsnm{Kempen}, \binits{E.G.M.}},
\bauthor{\bsnm{Verhaar}, \binits{B.J.}},
\bauthor{\bsnm{Kokkelmans}, \binits{S.J.J.M.F.}}:
\batitle{{Feshbach resonances with large background scattering length:
  Interplay with open-channel resonances}}.
\bjtitle{Phys. Rev. A}
\bvolume{70},
\bfpage{012701}
(\byear{2004}).
\doiurl{10.1103/PhysRevA.70.012701}
\end{barticle}
\endbibitem

\bibitem{nie01}
\begin{barticle}
\bauthor{\bsnm{Nielsen}, \binits{E.}},
\bauthor{\bsnm{Fedorov}, \binits{D.V.}},
\bauthor{\bsnm{Jensen}, \binits{A.S.}},
\bauthor{\bsnm{Garrido}, \binits{E.}}:
\batitle{{The three-body problem with short-range interactions}}.
\bjtitle{Phys. Rep.}
\bvolume{347},
\bfpage{373}
(\byear{2001}).
\doiurl{10.1016/S0370-1573(00)00107-1}
\end{barticle}
\endbibitem

\bibitem{gar18}
\begin{barticle}
\bauthor{\bsnm{Garrido}, \binits{E.}}:
\batitle{{Few-body techniques using coordinate space for bound and continuum
  states}}.
\bjtitle{Few-body Syst.}
\bvolume{59},
\bfpage{17}
(\byear{2018}).
\doiurl{10.1007/s00601-018-1354-2}
\end{barticle}
\endbibitem

\bibitem{fed01}
\begin{barticle}
\bauthor{\bsnm{Fedorov}, \binits{D.V.}},
\bauthor{\bsnm{Jensen}, \binits{A.S.}}:
\batitle{{Regularization of a three-body problem with zero-range potentials}}.
\bjtitle{J. Phys. A: Math. Gen.}
\bvolume{34},
\bfpage{6003}
(\byear{2001}).
\doiurl{10.1088/0305-4470/34/30/311}
\end{barticle}
\endbibitem

\bibitem{mar09}
\begin{barticle}
\bauthor{\bsnm{Marzok}, \binits{C.}},
\bauthor{\bsnm{Deh}, \binits{B.}},
\bauthor{\bsnm{Zimmermann}, \binits{C.}},
\bauthor{\bsnm{Courteille}, \binits{P.W.}},
\bauthor{\bsnm{Tiemann}, \binits{E.}},
\bauthor{\bsnm{Vanne}, \binits{Y.V.}},
\bauthor{\bsnm{Saenz}, \binits{A.}}:
\batitle{{Feshbach resonances in an ultracold $^7$Li and $^{87}$Rb mixture}}.
\bjtitle{Phys. Rev. A}
\bvolume{79},
\bfpage{012717}
(\byear{2009}).
\doiurl{10.1103/PhysRevA.79.012717}
\end{barticle}
\endbibitem

\bibitem{cui18}
\begin{barticle}
\bauthor{\bsnm{Cui}, \binits{Y.}},
\bauthor{\bsnm{Deng}, \binits{M.}},
\bauthor{\bsnm{You}, \binits{L.}},
\bauthor{\bsnm{Gao}, \binits{B.}},
\bauthor{\bsnm{Tey}, \binits{M.K.}}:
\batitle{{Broad Feshbach resonances in ultracold alkali-metal systems}}.
\bjtitle{Phys. Rev. A}
\bvolume{98},
\bfpage{042708}
(\byear{2018}).
\doiurl{10.1103/PhysRevA.98.042708}
\end{barticle}
\endbibitem

\bibitem{chr22}
\begin{barticle}
\bauthor{\bsnm{Christensen}, \binits{E.R.}},
\bauthor{\bsnm{Garrido}, \binits{E.}},
\bauthor{\bsnm{Jensen}, \binits{A.S.}}:
\batitle{{Two-body continuum states in noninteger geometry}}.
\bjtitle{Phys. Rev. A}
\bvolume{105},
\bfpage{033308}
(\byear{2022}).
\doiurl{10.1103/PhysRevA.105.033308}
\end{barticle}
\endbibitem

\bibitem{abr65}
\begin{botherref}
\oauthor{\bsnm{Abramowitz}, \binits{M.}},
\oauthor{\bsnm{Stegun}, \binits{I.A.}}:
{Handbook of Mathematical Functions}.
(Dover Publ., Inc., New York, 1965), p. 437
\end{botherref}
\endbibitem

\bibitem{dui04}
\begin{barticle}
\bauthor{\bsnm{Duine}, \binits{R.A.}},
\bauthor{\bsnm{Stoof}, \binits{H.T.C.}}:
\batitle{{Atom-molecule coherence in Bose gases}}.
\bjtitle{Phys. Rep.}
\bvolume{396},
\bfpage{115}
(\byear{2004}).
\doiurl{10.1016/j.physrep.2004.03.003}
\end{barticle}
\endbibitem

\end{thebibliography}


\end{document}